\documentclass[aps,superscriptaddress,nofootinbib]{revtex4-2}

\usepackage{epsfig,graphicx,amsfonts,amsmath,amssymb,amsthm,bbm}
\usepackage{bbold}
\usepackage[utf8]{inputenc}
\usepackage[T1]{fontenc}
\DeclareMathOperator{\Tr}{Tr}

\usepackage{booktabs}
\usepackage{float}

\RequirePackage{mathrsfs}
\usepackage{color}
\usepackage{epsf}
\usepackage{epstopdf}

\def\eq#1{{Eq.~(\ref{#1})}}
\newcommand{\Le}{\left(}
\newcommand{\Ra}{\right)}

\newcommand{\beq}{\begin{equation}}
	\newcommand{\eeq}{\end{equation}}
\newcommand{\beqar}{\begin{eqnarray}}
	\newcommand{\eeqar}{\end{eqnarray}}
\newcommand{\nn}{\nonumber}

\graphicspath{{./}}

\begin{document}

\title{Parton wave functions, density matrices and entropy in DGLAP evolution}
\author{S. Bondarenko}
\author{M. Ouchen}
\author{A. Prygarin}
\affiliation{Department of Physics, Ariel University, Ariel 40700, Israel}
\date{\today}

\begin{abstract}
Parton multiplicity entropy, proposed by Kharzeev and Levin as entanglement between the region probed in deep inelastic scattering and the proton's remainder, was compared with hadron multiplicity data. It is computed from parton densities, which do not fix the partons' quantum state. We show that a one-parton density matrix from Lipatov's wave functions determines two entropies for resolved modes with finite mean number and entropy. Its trace is the mean parton number, and normalization gives a sampled parton's entropy. A compatible thermal state, if it exists, maximizes entropy at fixed one-parton matrix. This maximum equals the actual entropy only for that thermal state and measures entanglement only for a specified subsystem of a pure state. The geometric multiplicity entropy follows if all emitted partons share one mode. An entropy computed from densities becomes such a maximum, an upper bound on the actual entropy, only for specified resolved modes. We also show that the state averaged over unrecorded emissions, assumed incoherent across scales, obeys probability-preserving Lindblad evolution on finitely many regulated modes with the resolution scale if emission and no-emission terms match. The occupations reproduce leading-order DGLAP, with virtual terms, at leading logarithmic accuracy as the cutoffs on momentum fractions are removed. The reduction holds for strongly ordered transverse momenta in light-cone gauge and dilute occupations, including those of newly resolved modes. Approximating momentum and helicity modes by independent thermal populations closes the occupation equation and gives this bound's evolution, with Bose and Pauli factors from the operator algebra. We propose testing this approximation at HERA or the LHC. The test compares the mode number inferred from normalized hadron factorial moments in a rapidity window with the window's transverse mode count.
\end{abstract}

\maketitle
\section{Introduction}\label{SecIntro}

Inclusive parton densities specify how quarks and gluons share the proton's momentum. Whether an entropy computed from these densities is an entropy of the proton state depends on the quantum information the densities discard. The relation between a measured multiplicity entropy and a quantum entropy requires a choice of subsystem and a specification of the parton state.
Kharzeev and Levin~\cite{KharzeevLevin} proposed relating deep inelastic scattering to partonic entanglement and computed the entropy of a geometric multiplicity distribution in a cascade without transverse dimensions. Hentschinski and Kutak~\cite{HentschinskiH1} compared a partonic entropy prescription with the entropy of charged-hadron multiplicity distributions measured by H1, while Hentschinski, Kutak and Straka~\cite{HKS2022} examined the dependence on the counted rapidity interval.
Tu, Kharzeev and Ullrich~\cite{TuKharzeevUllrich} tested the proposal with hadron multiplicity distributions measured by CMS in proton-proton collisions. Hentschinski, Kharzeev, Kutak and Tu~\cite{HentschinskiEE} compared the rapidity dependence of the evolved entropy with hadron entropies measured by H1 in rapidity windows. Further studies address entanglement and entropy in deep inelastic scattering and hadronic collisions~\cite{KutakEntropy,ZhangEnt,KharzeevLevinData,CaputaKrylov,BraunInclusive,ZhangKorepin,GotsmanLevin}.

Different choices of subsystem and retained information have been studied in the Color Glass Condensate description and in DGLAP evolution. Armesto et al.~\cite{ArmestoCGC} and Li and Kovner~\cite{LiKovner} obtained evolution of the form given by Lindblad~\cite{Lindblad} and Gorini, Kossakowski and Sudarshan~\cite{GKS} (Sec.~\ref{SecBC}). Kovner and Lublinsky~\cite{KovnerLublinsky} computed the entanglement entropy between soft gluons and valence charges in the Color Glass Condensate wave function. Duan, Akkaya, Kovner and Skokov~\cite{DuanIgnorance} defined an entropy of ignorance from a restricted set of measurements that fixes the diagonal of the density matrix in Fock space. Duan, Kovner and Skokov~\cite{DuanQuasiparticle} showed that the soft-gluon density matrix is thermal in a quasiparticle basis obtained from the gluon operators by a Bogoliubov transformation. The quasiparticle occupations include information from anomalous correlations, which pair two creation or two annihilation operators. 
Golec-Biernat~\cite{GolecBiernat} studied the relative entropy of the momentum-fraction distribution $x(\Sigma+g)$, where $\Sigma$ is the quark singlet, to the uniform distribution and showed that this relative entropy increases monotonically with the evolution scale for positive, decreasing momentum distributions.
Bondarenko and Prygarin~\cite{BondarenkoPrygarinInfoGeo} evaluated the Fisher information of the same normalized momentum density $P=x(\Sigma+g)$ along $\ln Q^2$ in DGLAP. It equals the momentum-weighted variance of $d\ln P/d\ln Q^2$ and twice the quadratic relative-entropy coefficient between neighboring scales. Its square root measures the speed of shape change, distinct from entropy at one scale.
These examples make the retained degrees of freedom and the information supplied about their state part of the definition of entropy.

The wave functions of Lipatov~\cite{LipDGLAP,LipatovSmallx}, also discussed by Ioffe, Fadin and Lipatov~\cite{IoffeFadinLipatov}, give the parton densities after the unobserved partons are summed over. Their complex amplitudes retain relative phases that disappear from an inclusive density. The light-front Fock expansion of Lepage and Brodsky~\cite{LepageBrodsky} and the review by Brodsky, Pauli and Pinsky~\cite{BrodskyPauliPinsky} provide the state and normalization conventions used here. A one-parton density matrix obtained from these amplitudes retains off-diagonal information while reproducing the densities on its diagonal.
Raufeisen and Brodsky~\cite{RaufeisenBrodsky} introduced a light-front quark density matrix obtained from light-front wave functions as the one-particle density matrix is obtained in nonrelativistic quantum mechanics. Its diagonal gives collinear quark densities, and its off-diagonal elements are related to generalized parton distributions.

Peschel and Eisler~\cite{PeschelEisler} reviewed reduced free-particle density matrices whose thermal form is fixed by their one-body correlations. Wolf, Giedke and Cirac~\cite{WGC} recalled that Gaussian states maximize the entropy at fixed covariance.
We apply these known properties to the resolved parton modes.

Dumitru and Kolbusz~\cite{DumitruKolbusz,DumitruKolbusz2} built reduced quark and gluon density matrices from proton light-front wave functions. Tracing out the quarks in the $|qqqg\rangle$ Fock state gives the density matrix of the gluon momentum fraction. Liu, Nowak and Zahed~\cite{LNZ} computed the entanglement entropy between fast and slow degrees of freedom in Mueller's dipole wave function~\cite{MuellerDipole}, separating the parton-number and within-sector contributions. They found a multiplicity distribution far from geometric in the double-logarithmic limit shared with DGLAP evolution.

The scale evolution of a parton density matrix was computed by Dumitru and Kolbusz~\cite{DumitruKolbuszDPDF} for one step, through the emission of one collinear gluon.
 Their calculation evolves the whole two-quark matrix in the momentum fractions, including its off-diagonal elements, for $\alpha_s\ln(Q^2/Q_0^2)\ll1$, and recovers the double-parton DGLAP equations on its diagonal. Duan, Kovner and Lublinsky~\cite{DuanCSS} obtained Lindblad-type evolution for fixed-$x$ observables, distinguishing the tagged rapidity from the evolution cutoff. Their frequency-ordered evolution of the light-cone wave function~\cite{DuanBORGsetup,DuanBORG} is compared with the collinear evolution in Section~\ref{SecBOccupations}.
In this paper we evolve the history-averaged parton state of the regulated modes continuously in the logarithm of the squared resolution scale. The one-parton matrix of this state gives the occupations and entropy functionals at each scale.

Probability conservation relates real emission to virtual corrections in parton evolution.
Lipatov~\cite{LipatovSmallx} used this relation in the wave-function description. 
Related evolution equations appear in the parton shower of Nagy and Soper~\cite{NagySoper}, the amplitude evolution of Forshaw, Holguin and Pl\"atzer~\cite{FHP}, the high-energy model of Chachamis, Hentschinski and Sabio Vera~\cite{ChachamisLindblad}, the spin evolution of Gu et al.~\cite{GuRG}, and the QED dephasing evolution of Rey~\cite{Rey2026}.
 Section~\ref{SecBC} compares their assumptions with the emission and no-emission description of Dalibard, Castin and M\o lmer~\cite{DalibardCastinMolmer} and Carmichael~\cite{Carmichael}.

The question here is how to apply these ideas to the parton Fock space while keeping the assumptions needed for an entropy calculation explicit. In this paper we define the one-parton density matrix from the wave functions and compare its two entropy functionals. We introduce the amplitude ansatz~\eqref{B4} because densities do not determine amplitudes. 
Section~\ref{SecBAmplitudes} fixes the boundary prescription at coincident creation scales.
  We separately average $|\psi\rangle\langle\psi|$ over unrecorded emission histories, treating different creation scales as incoherent, to obtain the mixed state of Section~\ref{SecBC}.  
We also treat splittings into more than two daughters, with probability conservation fixing the coefficients of the no-emission terms if the channels are added coherently and the splitting vertices are supplied (Sec.~\ref{SecBD}). Interference between channels with different numbers of daughters can raise or lower the parton number.

The ingredients of the approach are the known Gaussian entropy maximum~\cite{PeschelEisler,WGC}, reduced parton matrices~\cite{RaufeisenBrodsky,DumitruKolbusz,DumitruKolbuszDPDF}, and emission and no-emission evolution~\cite{NagySoper,FHP}.
 We apply these ingredients to collinear splitting operators on parton Fock space.
The occupation equation derived here includes Pauli blocking and Bose enhancement of final-state partons. Mangano, Miele and Migliore~\cite{ManganoMieleMigliore} inserted such factors into the leading-order DGLAP equations by analogy with the Boltzmann collision integral. Their occupation was the parton distribution divided by a weight function independent of the scale.
Stimulated emission is also present in the frequency-ordered cascade~\cite{DuanBORG}.

The paper is organized as follows.
Section~\ref{SecA} defines the one-parton matrix and its entropies. Section~\ref{SecB} gives the amplitude ansatz, the emission-history evolution and its DGLAP reduction. Section~\ref{SecResolution} examines resolution and occupancy scaling, and Section~\ref{SecConclusions} discusses the results. Appendices~\ref{AppA} through \ref{AppD} contain the normalization and derivations.

We show that Lipatov's parton wave functions determine the entropy of a sampled parton after normalization and the maximum state entropy at fixed one-parton density matrix, for specified resolved modes with finite traces and entropy if a compatible thermal state exists.
 The trace is the mean parton number. 
At leading twist the diagonal gives collinear densities and light-cone-gauge densities in transverse momentum, while the off-diagonal helicity elements retain phases at equal momentum.
 The maximum in Eq.~\eqref{B28} equals the actual state entropy only for that thermal Gaussian state and describes entanglement only for a specified subsystem of a pure joint state. Its one-mode case is the known geometric multiplicity entropy if all emitted partons occupy that mode. We also show that the parton state averaged over unrecorded emission histories, treated as incoherent at different scales, obeys the probability-preserving Lindblad equation~\eqref{B12} in $t=\ln Q^2$ on a finite set of regulated modes when the no-emission term is matched to the emission probability as in Eq.~\eqref{B11}. 
The occupations then reproduce leading-order DGLAP, including its virtual terms, at leading logarithmic accuracy for dilute occupations, including those of newly resolved modes, and strongly ordered transverse momenta in light-cone gauge, as the cutoffs on momentum fractions are removed.
The longitudinal momentum of the resolved modes is conserved for any parton state and every truncation that retains momentum-conserving splitting vertices with their matched virtual terms.
 Approximating the momentum and helicity modes by independent thermal populations, a Gaussian state with coherence between distinct modes neglected, closes the occupation equation with Bose and Pauli factors from the operator algebra and determines the mode-entropy production. These results specify which entropies the one-body data determine and the statistical assumptions needed to follow the mode entropy during evolution.

\section{Parton wave functions and one-parton information}\label{SecA}

\subsection{Wave functions and parton densities}\label{SecAWavefunctions}

A parton density sums over the unobserved partons and retains the probability for the observed momentum. The wave function also retains relative phases. We use light-cone gauge, $A^{+}=0$, and the light-front Fock basis, with no gauge link in the one-parton operator below.

The scale dependence of the collinear parton distributions is described by the equation of Gribov and Lipatov~\cite{LipDGLAP}, Dokshitzer~\cite{Dokshitzer}, and Altarelli and Parisi~\cite{AltarelliParisi},
\beq\label{AA1}
Q^2\,\frac{\partial f_i(x,Q^2)}{\partial Q^2}\,=\, \sum_j \int_x^1 \frac{dz}{z}\, P_{ij}(z,\alpha_s)\,f_j(x/z,Q^2)
\eeq
where $f_i(x,Q^2)$ is the density of species $i$ at the hard scale $Q^2$. The splitting function $P_{ij}$ describes $j\to i+k$. Its argument is the fraction $z=x_i^{\rm observed}/x_j^{\rm parent}$, with $x\leq z\leq1$, so the parent in Eq.~\eqref{AA1} has longitudinal fraction $x/z$. Equivalently, $z=k^+_{\rm observed}/k^+_{\rm parent}$.

Lipatov's parton wave functions~\cite{LipDGLAP,LipatovSmallx}, see also Section 7.3 of Ref.~\cite{IoffeFadinLipatov}, give the amplitude for each Fock configuration,
\beq\label{AA2}
|h\rangle \,=\,\sum_n \int [d\mu_n]\; \psi_n(\{x_i, k_{\perp,i}\})\; |n;\{x_i, k_{\perp,i}\}\rangle,
\eeq
and the constrained measure is
\beq\label{AA3}
[d\mu_n] = 16\pi^{3}\,\delta\Bigl(1-\sum_{i=1}^{n}x_{i}\Bigr)\,\delta^{(2)}\Bigl(\sum_{i=1}^{n}k_{\perp,i}\Bigr)\,\prod_{i=1}^{n}\dfrac{dx_i\, 
d^2 k_{\perp,i}}{16\pi^3}\,,
\eeq
where the measure includes the flavor, color and spin sums. The factor $16\pi^3$ compensates the constrained momentum, in the conventions of Lepage and Brodsky~\cite{LepageBrodsky} and Brodsky, Pauli and Pinsky~\cite{BrodskyPauliPinsky}. A state of fixed hadron momentum is not normalizable. We assume a normalizable hadron packet and use its canonical Fock amplitudes for the one-parton reduction. 
The packet has the same sector probabilities as the intrinsic amplitudes in Eq.~\eqref{AA2}. Appendix~\ref{AppAFock} gives their normalization and the Fock symmetry factors.
 The intrinsic amplitudes are normalized by
\beq\label{AA4}
\sum_n \int [d\mu_n]\;|\psi_n(\{x_i, k_{\perp,i}\})|^2 = 1.
\eeq
The pure hadron state has density operator
\beq\label{AB1}
\hat\rho_{\rm pure}\, = \,|h\rangle\langle h|.
\eeq
Its von Neumann entropy vanishes. We use natural logarithms and $k_B=1$ throughout.

The observed parton has label and measure
\beq\label{AB4}
\xi \;\equiv\; (x,\, k_\perp,\, \text{flavor},\, \text{color},\, \text{spin}),\qquad
\int d\xi \;\equiv\; \sum_{\rm f,c,s}\int \frac{dx\, d^2 k_\perp}{16\pi^3},
\eeq
and the creation and annihilation operators satisfy
\beq\label{AB5}
[\,a(\xi),\,a^\dagger(\xi')\,]_{\mp} = (16\pi^3)\,\delta_{\rm fcs}\,
\delta(x-x')\,\delta^2(k_\perp-k_\perp')\,,
\eeq
where $[A,B]_{\mp}=AB\mp BA$. The upper sign is the gluon commutator and the lower sign the quark anticommutator. The species Kronecker symbol prevents an exchange of different species from contributing to the identical-particle permutation sum.

The one-parton reduced density matrix is the number-conserving correlation function
\beq\label{AB6}
\hat\rho_1(\xi;\xi')\;=\; \langle h|\, a^\dagger(\xi')\, a(\xi)\, |h\rangle 
\eeq
whose spectator trace gives the following expression in terms of the canonical packet amplitudes, denoted by $\psi_n$ in this expression,
\beq\label{AB7}
\hat\rho_1(\xi;\xi')\, =\, \sum_{n\ge 1} n \int \prod_{i=2}^{n}d\xi_i\;\psi_n(\xi,\{\xi_i\})\,\psi_n^{*}(\xi',\{\xi_i\})\,,
\eeq
where the integration includes all spectator quantum numbers. At fixed species, $n$ counts that species in the sector. Summing the species gives the total multiplicity. The factor $n$ counts the possible observed partons and follows from the Fock symmetry factors, as derived in Appendix~\ref{AppAFock}. The bilinear $a^\dagger a$ does not probe matrix elements between different total-parton-number sectors. Their absence from Eq.~\eqref{AB7} is a property of this observable and does not require the full state to be diagonal in parton number.

The kernel is Hermitian and positive semidefinite, as shown in Appendix~\ref{AppAFock}. Its trace counts partons,
\beq\label{AB9}
\Tr\hat\rho_1\, =\, \int d\xi\; \hat\rho_1(\xi;\xi)\,=\, \langle h|\, \hat N\, |h\rangle\,=\, \langle N \rangle,
\eeq
where the resolved modes are specified before taking the trace and $0<\langle N\rangle<\infty$ is required. In particular, the integral of a physical gluon number density diverges at small $x$ if $xG(x,Q^2)$ is constant or rises as $x$ decreases. A longitudinal cutoff $x\geq x_{\rm min}$ is then required. Finiteness of each entropy used below is a separate assumption.

Longitudinal momentum conservation sets $\langle a(\xi)a(\xi')\rangle=0$ for $x,x'>0$ in a state of definite total $P^+$, or any state commuting with the total longitudinal momentum operator $\hat P^+$. The product $a(\xi)a(\xi')$ lowers that momentum. The momentum-conserving evolution of Section~\ref{SecB} cannot generate anomalous covariance from such a state. A packet of finite width, or another state that does not commute with $\hat P^+$, can have coherence between sectors differing by two partons and nonzero $\langle aa\rangle$. These anomalous correlators are absent from Eq.~\eqref{AB6}. A Gaussian state with them is of Bogoliubov type and requires the additional covariance in Eq.~\eqref{B29} to determine its entropy.

The diagonal of Eq.~\eqref{AB6} is the light-cone-gauge parton number density in $x$ and $k_\perp$. The gauge-invariant transverse-momentum correlator also contains a transverse gauge link at light-cone infinity. This link cancels in the integral over all transverse momenta, as Belitsky, Ji and Yuan~\cite{BelitskyJiYuan} showed. The integral up to $|k_\perp|<Q$ in Eq.~\eqref{AB16} then gives the leading-twist collinear densities at leading logarithmic accuracy. The unintegrated diagonal coincides with the time-reversal-even transverse-momentum distribution with final-state interactions neglected, as in the tree-level example of Section~\ref{SecAHelicity}. It is not identified with that distribution beyond tree level.
 A projector $P_x$ onto a longitudinal slice $[x,x+dx]$ defines $\hat\rho_1^{(x)}=P_x\hat\rho_1P_x$, with
\beq\label{AB16}
\Tr\hat\rho_1^{(x)}\,=\,dx\sum_{\rm f,c,s}\int\frac{d^{2}k_{\perp}}{16\pi^{3}}\;
n\bigl(x,k_{\perp},{\rm f},{\rm c},{\rm s};Q^{2}\bigr)\,=\,f(x,Q^{2})\,dx\,,
\eeq
where $n(\xi)=\hat\rho_1(\xi;\xi)$ is the diagonal kernel, with dimension $M^{-2}$. The integral runs over the resolved modes with $|k_\perp|<Q$, and the argument $Q^2$ specifies this resolution.
 The flavor-summed transverse-momentum-dependent density is
\beq\label{AB16001}
f(x,k_{\perp},Q^{2})=\sum_{\rm f,c,s}n(\xi)/16\pi^{3}
\eeq
and $f(x,Q^2)=\sum_i f_i(x,Q^2)$ is its transverse integral. 
The restriction $|k_\perp|<Q$ follows Lepage and Brodsky~\cite{LepageBrodsky}. The density of species $i$ at scale $Q$ further restricts the sum in Eq.~\eqref{AB16} to that species.
 Its spectator representation retains the parton-counting factor of Eq.~\eqref{AB7}. 

The diagonal kernel gives an occupation per transverse mode per unit $x$, $n(\xi)/(4\pi S_\perp)$, with $S_\perp$ the hadron's transverse area. The occupation per mode follows only after the longitudinal cell and its number of retained modes are specified, as in Section~\ref{SecResolution}.

\subsection{Entropy of a sampled parton}\label{SecAC}

We diagonalize the one-parton matrix,
\beq\label{AB10}
\int d\xi'\;\hat\rho_1(\xi;\xi')\,u_{\lambda}(\xi')\,=\,n_{\lambda}\,u_{\lambda}(\xi)\,,\qquad
\int d\xi\;u_{\lambda}^{*}(\xi)\,u_{\mu}(\xi)\,=\,\delta_{\lambda\mu}\,,
\eeq
and give the spectral representation
\beq\label{AB11}
\hat\rho_1\,=\,\sum_{\lambda}n_{\lambda}\,|\lambda\rangle\langle\lambda|\,,\qquad
n_{\lambda}\,\geq\,0\,,\qquad \sum_{\lambda}n_{\lambda}\,=\,\langle N\rangle\,.
\eeq
The functions $u_\lambda$ are the natural orbitals and $n_\lambda$ their dimensionless occupations. Quark modes obey $0\leq n_\lambda\leq1$, while gluon occupations have no upper bound.

A randomly sampled parton is described by the trace-one matrix
\beq\label{AB14}
\varrho\,\equiv\,\frac{\hat\rho_1}{\Tr\hat\rho_1}\,=\,\frac{\hat\rho_1}{\langle N\rangle}\,,\qquad \Tr\varrho\,=\,1\,,
\eeq
with eigenvalues $p_\lambda=n_\lambda/\langle N\rangle$ and $\sum_\lambda p_\lambda=1$. 
The entropy of a randomly sampled parton is
\beq\label{AB15}
S_{1}\,=\,-\,\Tr\bigl[\,\varrho\,\ln\varrho\,\bigr]\,=\,-\sum_{\lambda}p_{\lambda}\ln p_{\lambda}\,.
\eeq
It measures the spread of the sampled parton over the natural orbitals. It vanishes for one occupied orbital and is bounded by $\ln M$ for $M$ occupied orbitals. The unnormalized expression $-\Tr[\hat\rho_1\ln\hat\rho_1]$ is a different quantity and can be negative, as shown in Appendix~\ref{AppAEntropy}.

The same entropy can be expressed in the physical momentum basis. We use finite longitudinal cells $i$ and transverse cells $j$, keeping flavor, color and spin as internal labels. The normalized weight of longitudinal cell $i$ is $w_i=\int_{{\rm cell}\,i}dx\,f(x,Q^2)/\langle N\rangle$. The probability $p_{j|i}$ is the transverse-cell probability conditioned on cell $i$, and $W_{ij}=w_i p_{j|i}$ is the joint probability. 
The normalized longitudinal block is $\varrho_{x_i}$. The normalized block $\varrho_{j,x_i}$ of cell $(i,j)$ retains any momentum dependence inside the cell. It reduces to the internal state of flavor, color and spin for a cell containing one momentum mode, as in Section~\ref{SecResolution}.
 For a narrow slice, $p_{j|x}$ is the integral over transverse cell $j$ of $p(k_\perp|x)=f(x,k_\perp,Q^2)/f(x,Q^2)$, and $\varrho_x=\hat\rho_1^{(x)}/\Tr\hat\rho_1^{(x)}$.

The map $\Delta$ removes coherence between transverse cells. The entropy increase from this dephasing is
\beq\label{AB17002}
C_{\rm rel}(\varrho_{x})\,=\,S(\Delta\varrho_{x})-S(\varrho_{x})\geq0
\eeq
and the full one-parton entropy is
\beq\label{AB17003}
S_{1}\,=\,-\sum_{ij}W_{ij}\,\ln W_{ij}
\;+\;\sum_{ij}W_{ij}\,S(\varrho_{j,x_i})\;-\;\sum_i w_i\,C_{\rm rel}(\varrho_{x_i})\,,
\eeq
provided $\hat\rho_1$ is block diagonal in the longitudinal cells. Otherwise an additional longitudinal coherence term $C_x\geq0$ is subtracted. The first term is the Shannon entropy of the joint cell probabilities. 
The block entropy $S(\varrho_{j,x_i})$ enters the second term of Eq.~\eqref{AB17003} with a sign opposite to the transverse coherence. The coherence $C_{\rm rel}$ vanishes for a state block diagonal in the momentum cells. The eigenvalues are then $p_\lambda=W_{ij}q_a$, with $q_a$ the eigenvalues of $\varrho_{j,x_i}$. The block entropy and the Shannon term give Eq.~\eqref{AB15} for that state.
 In continuum notation, the Shannon term becomes a differential entropy only up to the logarithm of the cell volume, which is not part of $C_{\rm rel}$. A hadron of definite momentum has $\hat\rho_1$ diagonal in $(x,k_\perp)$, as in the limit described in Appendix~\ref{AppAFock}. The terms $C_{\rm rel}$ and $C_x$ then vanish and arise only from the finite packet width. Off-diagonal internal elements at equal momentum can remain, such as the helicity elements of Section~\ref{SecAHelicity}. Appendix~\ref{AppAEntropy} gives the decomposition at each conditioning level.

\subsection{Entropy assigned to mode occupations}\label{SecAMode}

The one-parton matrix can also specify the largest entropy of a state on the Fock space of the resolved modes. This entropy measures the uncertainty about parton numbers in the modes when only that matrix is known.
 The maximizing state is thermal Gaussian,
\beq\label{AB18}
\hat\varrho_{\rm mode}\,=\,\frac{e^{-\hat K}}{Z}\,,\qquad
\hat K\,=\,\sum_{\lambda}\epsilon_{\lambda}\,b_{\lambda}^{\dagger}b_{\lambda}\,.
\eeq
Here $b_\lambda$ annihilates a natural orbital and $Z$ normalizes the state. Thermal Gaussian means that Wick factorization determines higher correlators from the normal two-point function, with vanishing one-point functions and anomalous correlators $\langle b_\lambda b_\mu\rangle$. The parameters $\epsilon_\lambda$, often called modular energies, are fixed by
\beq\label{AB19}
n_{\lambda}\,=\,\frac{1}{e^{\epsilon_{\lambda}}\mp 1}\,;\qquad \Longrightarrow\;\qquad
\epsilon_{\lambda}\,=\,\ln\Le 1+\frac{1}{n_{\lambda}}\Ra\;\;(\text{gluons}),\qquad
\epsilon_{\lambda}\,=\,\ln\frac{1-n_{\lambda}}{n_{\lambda}}\;\;(\text{quarks}),
\eeq
as used in the reduced free-particle density matrices obtained from correlation functions by Peschel~\cite{Peschel2003} and the mode entropies given by Peschel and Eisler~\cite{PeschelEisler}. In the present notation the gluon functional is
\beq\label{AB20}
S_{\rm mode}\,=\,\sum_{\lambda}\Bigl[(1+n_{\lambda})\ln(1+n_{\lambda})\,-\,n_{\lambda}\ln n_{\lambda}\Bigr]\;\qquad(\text{gluons}),
\eeq
and the quark functional is
\beq\label{AB21}
S_{\rm mode}\,=\,-\sum_{\lambda}\Bigl[n_{\lambda}\ln n_{\lambda}\,+\,(1-n_{\lambda})\ln(1-n_{\lambda})\Bigr]\qquad(\text{quarks}).
\eeq
We denote the per-mode functions by
\beq\label{AB21002}
g_{B}(n)\,=\,(1+n)\ln(1+n)\,-\,n\ln n\,,\qquad
g_{F}(n)\,=\,-\,\bigl[\,n\ln n\,+\,(1-n)\ln(1-n)\,\bigr]\,,
\eeq
so that $S_{\rm mode}=\sum_\lambda g_B(n_\lambda)$ for gluons and $\sum_\lambda g_F(n_\lambda)$ for quarks. At fixed gluon species we also write $g=g_B$.

The maximum-entropy property requires that the thermal state with the specified normal covariance exist and that $S_{\rm mode}$ be finite. It is the quadratic Gibbs case of the entropy variational principle discussed by Wehrl~\cite{Wehrl}, Holevo, Sohma and Hirota~\cite{HSH}, and Wolf, Giedke and Cirac~\cite{WGC}. Any Fock-space state $\hat\rho$ with the same one-parton matrix has von Neumann entropy $S_{\rm vN}=-\Tr\hat\rho\ln\hat\rho$ satisfying
\beq\label{B28}
S_{\rm vN}\;\leq\;\sum_{\lambda}g\bigl(n_{\lambda}\bigr)\,=\,S_{\rm mode}\,,
\eeq
where the difference $S_{\rm mode}-S_{\rm vN}(\hat\rho)$ is the relative entropy of $\hat\rho$ to $\hat\varrho_{\rm mode}$. Equality holds only when $\hat\rho=\hat\varrho_{\rm mode}$. The same bound holds for quark modes with $g_F$ in place of $g$, by the same argument applied to Eq.~\eqref{AB21}. An entanglement interpretation also requires the resolved modes to be a specified subsystem of a pure joint state. The pure hadron of Eq.~\eqref{AB1} has zero entropy even though its one-parton occupations can give a positive $S_{\rm mode}$.
Alicki~\cite{Alicki2019} defined the entropy of a bosonic field's single-particle density matrix and averaged field as the largest von Neumann entropy compatible with both, obtained the functional of Eq.~\eqref{AB20} for a vanishing averaged field, and noted that this entropy vanishes only for coherent states.

The thermal state of Eq.~\eqref{AB18} is the maximum-entropy assignment for the one-parton data. It does not describe the pure hadron, because momentum constraints correlate its modes and the color-singlet constraint violates Gaussian factorization.
 Section~\ref{SecB} uses the thermal form for color-averaged occupations of the state averaged over unrecorded emission histories, with coherence between distinct momentum and helicity modes neglected. 
The assumed thermal Gaussian state has every connected cumulant $\hat C_m$ of order $m\geq2$ equal to zero, as in Eq.~\eqref{B15001}. These cumulants are the connected parts of the reduced operators $\hat\rho_m$ in Appendix~\ref{AppD}. The converse does not hold because these cumulants are built from number-conserving operators and do not detect coherence between sectors of different parton number.
 
The entropy in Eq.~\eqref{B29} is strictly lower than $S_{\rm mode}$ at the same normal covariance when the Gaussian state has nonzero $\langle aa\rangle$. Such covariance requires a packet of finite width or another state that does not commute with $\hat P^+$, as explained in Section~\ref{SecAWavefunctions}.

The gluon entropy has the limits
\beq\label{AB22}
S_{\rm mode}\,\simeq\,n_{\lambda}\Bigl(1+\ln\frac{1}{n_{\lambda}}\Bigr)\;\;(n_{\lambda}\ll1)\,,\qquad
S_{\rm mode}\,\simeq\,1+\ln n_{\lambda}\;\;(n_{\lambda}\gg1)\,,
\eeq
per mode.

\subsection{Phase information in a quark splitting}\label{SecAHelicity}

One massless, unpolarized quark parent gives a two-helicity gluon block after the tree-level splitting $q\to qg$. In this example $z$ is the outgoing quark fraction, the gluon fraction is $1-z$, and $\varphi_k$ is the azimuth of the relative transverse momentum $k_\perp$ of Appendix~\ref{AppC}, carried by the quark. The gluon azimuth differs by $\pi$ and gives the same $e^{-2i\varphi_k}$. The phase convention and helicity amplitudes are given in Appendix~\ref{AppC}. The common factor is $g_s^2 C_F\Delta_1\Delta_2/(4\pi^2 k_\perp^4)$, with $g_s^2=4\pi\alpha_s$ and $C_F=(N_c^2-1)/(2N_c)$. Here $\Delta_1$ is the quark no-emission probability before the splitting, and $\Delta_2$ is the product of the quark and gluon no-emission probabilities afterward, as in Eq.~\eqref{eq:appb-psi2}. Dividing by this factor gives the block
\beq\label{eq:appb-rhohel}
\hat\rho_{1}^{++}\,=\,\hat\rho_{1}^{--}\,=\,\frac{k_{\perp}^{2}}{4}\,\frac{1+z^{2}}{1-z}\,,\qquad
\hat\rho_{1}^{+-}\,=\,-\,\frac{k_{\perp}^{2}}{2}\,\frac{z}{1-z}\;e^{-2i\varphi_{k}}\,,
\eeq
with $\hat\rho_1^{-+}$ the conjugate. The trace gives the splitting probability, proportional to $\hat P^{(0)}_{qq}(z)=C_F(1+z^2)/(1-z)$, with the hat marking the real kernel without its $\delta(1-z)$ endpoint term, as defined in Section~\ref{SecB}. The off-diagonal element retains the phase $e^{-2i\varphi_k}$ at equal momentum and vanishes upon azimuthal averaging. Mulders and Rodrigues~\cite{MuldersRodrigues} obtained this angular dependence for the helicity matrix of a linearly polarized gluon distribution. The comparison concerns the tensor and azimuthal form, without identifying the absolute normalization with a physical transverse-momentum distribution.

The degree of linear polarization is
\beq\label{eq:appb-pol2}
\frac{\bigl|\hat\rho_{1}^{+-}\bigr|}{\hat\rho_{1}^{++}}\,=\,\frac{2z}{1+z^{2}}\,,
\eeq
which vanishes as the gluon takes all the parent momentum and tends to unity as $z\to1$. In the latter limit the positivity bound is saturated, the block has rank one, and its internal entropy vanishes. The trace-normalized eigenvalues and the entropy increase from helicity dephasing are given in Eq.~\eqref{eq:appb-lam}. This helicity coherence is distinct from the transverse-cell coherence in Eq.~\eqref{AB17002}. The helicity block contributes its internal entropy to Eq.~\eqref{AB17003}, and retaining its off-diagonal element changes that entropy without changing the inclusive density.

The gauge convention limits the distributions represented by this example. 
A transverse gauge link at light-cone infinity remains for generic residual boundary conditions even when the lightlike link becomes unity in $A^+=0$ gauge~\cite{BelitskyJiYuan}.
 The transverse link vanishes for the advanced boundary condition, but the final-state phase is then carried by the light-cone wave functions~\cite{BelitskyJiYuan}. The principal-value prescription used here for the pole $1/(n\cdot k)$, as specified in Appendix~\ref{AppB}, introduces no absorptive final-state phase in the light-cone wave functions~\cite{BelitskyJiYuan}. The final-state phase then resides only in the transverse link, which Eq.~\eqref{AB6} does not carry. The correlator of Eq.~\eqref{AB6}, which omits the transverse link, therefore lacks the phase responsible for the time-reversal-odd distributions of Sivers~\cite{Sivers} and of Boer and Mulders~\cite{BoerMulders}. 
The correspondence with transverse-momentum distributions is the tree-level one shown in this example. A physical distribution also needs the transverse link, a rapidity regulator and a soft factor, as discussed by Collins~\cite{Collins}.

\subsection{One occupied mode and many occupied modes}\label{SecAOneMode}

Kharzeev and Levin~\cite{KharzeevLevin} obtained the shifted geometric multiplicity distribution
\beq\label{AB23}
p_{n}(Y)\,=\,e^{-\omega Y}\Le 1-e^{-\omega Y}\Ra^{n-1}\,,\qquad n\geq1\,,\qquad \bar n\,=\,e^{\omega Y}\,,
\eeq
for a cascade of growth rate $\omega$ over rapidity $Y$. The initial parton is always present, so the emitted multiplicity has mean $\bar n-1$. The Shannon entropy $S_{KL}=-\sum_{n\geq1}p_n\ln p_n$ is
\beq\label{AB26001}
S_{KL}\,=\,g_{B}\bigl(\bar n-1\bigr)\,,\qquad
g_{B}(n)\;\hbox{of}\;\eq{AB21002}\,,
\eeq
a closed form obtained by Kharzeev and Levin~\cite{KharzeevLevin}. Kutak and Prasza{\l}owicz~\cite{KutakPraszalowicz} wrote it as $\bar n\ln\bar n-(\bar n-1)\ln(\bar n-1)$. Its large-occupancy constant, the additive unit of Eq.~\eqref{AB27}, appears in Hentschinski, Kutak and Straka~\cite{HKS2022} and in Hentschinski, Kharzeev, Kutak and Tu~\cite{HentschinskiEE}.
The geometric law has the largest Shannon entropy among distributions with the same mean, so for one bosonic mode the thermal state maximizes entropy at fixed mean occupation, the one-mode case of Eq.~\eqref{B28}. Ouchen and Prygarin~\cite{OuchenPrygarinAGK} noted the memorylessness of the geometric law of Kharzeev and Levin and described its entropy as maximal for the same parameters.
Equation~\eqref{AB26001} identifies this known entropy with the one-mode case of $S_{\rm mode}$ if all emitted partons occupy that mode. Equality of these entropy functions does not identify the dynamics of the cascade with a one-mode parton splitting, whose rates differ as shown in Appendix~\ref{AppAOneMode}.

The comparison with a gluon density uses the identification of Kharzeev and Levin~\cite{KharzeevLevin}, $\bar n=xG(x,Q^2)$, for the mean cascade multiplicity, with which they obtained $S\simeq\ln[xG]$ at small $x$. The large-occupancy form is then
\beq\label{AB27}
S(x,Q^{2})\,\simeq\,\ln\bigl[x\,G(x,Q^{2})\bigr]\,+\,1\,,
\eeq
where the additive unit comes from the expansion of the known geometric entropy and lies beyond the accuracy of the logarithmic form.

The shift from $\bar n$ to $\bar n-1$ changes the entropy by a relative amount $1/(xG\ln xG)$, also beyond that accuracy, so either argument can be used in Eq.~\eqref{AB27}.
The scaled multiplicity is $\zeta=n/\langle n\rangle$, and $f_s(\zeta)$ is the relative deviation of the scaled distribution from $e^{-\zeta}$. Ouchen, Prygarin and Sandeu~\cite{OuchenPrygarinSandeu} identified the unit with the entropy of $e^{-\zeta}$. The normalized unit-mean scaling function is $\langle n\rangle P_n=e^{-\zeta}[1+f_s(\zeta)]$ in the continuum limit of Koba, Nielsen and Olesen (KNO) scaling~\cite{KNO}. The entropy is $S=\ln\langle n\rangle+1-\tfrac12\int_0^\infty d\zeta\,e^{-\zeta}f_s(\zeta)^2+O(f_s^3)$ for charged-hadron multiplicities, not the parton count $xG$~\cite{OuchenPrygarinSandeu}. Normalization and unit mean cancel the linear term, so deviations lower the entropy at second order.

\begin{figure}[htbp]
\centering
\includegraphics[width=\textwidth]{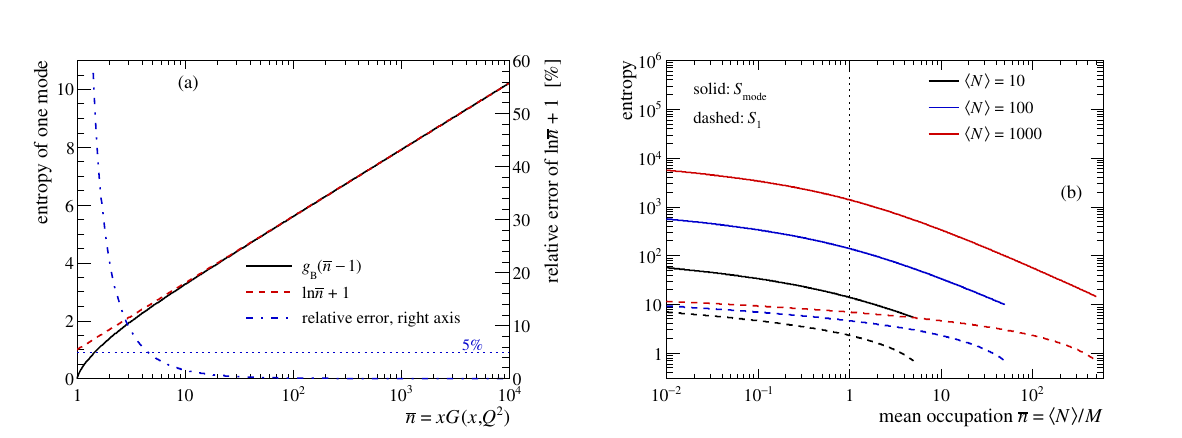}
\caption{(a) One-mode entropy $g_B(\bar n-1)$, \eq{AB26001} (solid black), and its large-occupancy form $1+\ln\bar n$, \eq{AB27} (dashed red), versus mean multiplicity $\bar n=xG$. The dash-dotted blue curve gives the latter's relative error against the former on the right axis, with a dotted blue five-percent line. (b) $S_{\rm mode}=Mg_B(\bar n)$ (solid) and $S_1=\ln M$ (dashed) for $M\geq2$ equally occupied modes at fixed total parton number $\langle N\rangle=10,10^2,10^3$ (black, blue, red), versus occupation $\bar n=\langle N\rangle/M$. The vertical dotted line marks $\bar n=1$.}
\label{fig:entropies}
\end{figure}

 This comparison treats the rapidity cascade at fixed resolution, so the identification of its multiplicity with $xG$ is an input. 
The large-occupancy form $1+\ln\bar n$ of Eq.~\eqref{AB27} overestimates $g_B(\bar n-1)$ by 47 percent at $\bar n=1.5$ and 1.6 percent at $\bar n=10$.
 Its relative error falls below five percent at $\bar n=4.53$, as shown in Fig.~\ref{fig:entropies}(a).

The sampled-parton entropy is zero for a single occupied orbital. A nonzero multiplicity entropy in that case measures occupation fluctuations, which are absent from the trace-one one-parton matrix. The occupation of each of $M$ equally occupied modes is $n_\lambda=\bar n\equiv\langle N\rangle/M$, and $S_1=\ln M$. Here $\bar n$ denotes the occupation per mode, distinct from the mean cascade multiplicity in Eq.~\eqref{AB23}. The large-occupancy relation is
\beq\label{AB28001}
S_{\rm mode}\,=\,M\,\Bigl[\,1\,+\,\ln\langle N\rangle\,-\,S_{1}\,\Bigr]\,,\qquad M\,=\,e^{S_{1}}\,,
\qquad(\hbox{$M$ equally occupied modes},\;\bar n\gg1)\,.
\eeq
The prefactor is the number of modes, and the relation is not a proportionality between the two entropies. In the dilute limit, $S_{\rm mode}\simeq\langle N\rangle[1+\ln(1/\bar n)]$. Appendix~\ref{AppAOneMode} gives the corresponding ratios.

Figure~\ref{fig:entropies}(b) holds the total parton number fixed and varies the number of equally occupied modes. The entropy $S_1=\ln M$ changes only logarithmically, while $S_{\rm mode}$ changes by orders of magnitude and falls as $\bar n$ grows. Concentrating the same partons into fewer modes lowers their mode entropy. The two functionals cannot be interchanged by a normalization factor. They answer different questions, the spread of one sampled parton over the natural orbitals and the largest Fock-space entropy compatible with the occupations. 

\section{Evolution of the parton state}\label{SecB}\label{SecEvolution}

An increase in transverse resolution allows a parent parton to split into two resolved partons. The splitting amplitude retains the helicity phases described in Section~\ref{SecA}, while its squared modulus gives the real-emission part of the DGLAP kernel.

\subsection{Splitting amplitudes and scale ordering}\label{SecBAmplitudes}

The coupling convention for the leading-order kernels is
\beq\label{B1002}
P_{ij}(z,\alpha_{s})\,=\,\frac{\alpha_{s}(Q^{2})}{2\pi}\,P^{(0)}_{ij}(z)\,+\,O(\alpha_{s}^{2})\,,
\eeq
where $C_F=(N_c^2-1)/2N_c$, $C_A=N_c$ and $T_F=1/2$ per quark flavor. The real kernels $\hat P^{(0)}_{ij}$ contain no virtual $\delta(1-z)$ term or plus prescription. Their relation to the splitting amplitudes is
\beq\label{B2}
\sum_{\rm spins,\, colors}\bigl|\mathcal V_{ij}(z, k_\perp)\bigr|^2\;=\; \frac{1}{k_\perp^{2}}\,
\frac{\alpha_{s}(k_{\perp}^{2})}{2\pi}\,\hat P^{(0)}_{ij}(z)\,,
\eeq
where the sum includes the final spins and colors and an average over those of the parent. The leading-order real kernel at $z<1$ is the square of the tree-level splitting wave function, as used by Lappi, Paatelainen and Sepp\"al\"a~\cite{LappiNLOLCPT}. Here $z$ is the observed parton's longitudinal momentum fraction relative to the parent, and $k_\perp$ is its transverse momentum relative to the parent. In the channel $j\to ik$, the parent is $j$ and the observed daughter is $i$. Thus $z$ is the quark fraction in Eq.~\eqref{B3}. The regularized kernels are listed in Appendix~\ref{AppB}.

The light-cone gauge vertex for $q\to qg$, normalized to Eq.~\eqref{B2}, is
\beq\label{B3}
\mathcal V^{\lambda'\lambda_{g}}_{\lambda}(z,k_\perp)\;=\;g_{s}\,T^{a}\,
\frac{\bigl(\epsilon^{*}_{\lambda_{g}}\!\cdot k_\perp\bigr)}{k_\perp^{2}}\;
C^{\lambda'\lambda_{g}}_{\lambda}(z)\,,\qquad
\sum_{\lambda',\lambda_{g}}\bigl|C^{\lambda'\lambda_{g}}_{\lambda}(z)\bigr|^{2}\;\propto\;\frac{1+z^{2}}{1-z}\,.
\eeq
Here $\alpha_s=g_s^2/4\pi$, $T^a$ is a fundamental $SU(N_c)$ generator, and $a$ is the emitted gluon's color. The coefficients $C^{\lambda'\lambda_g}_{\lambda}$ are fixed by the Dirac matrix elements in Tables~II and~III of Lepage and Brodsky~\cite{LepageBrodsky} for their helicity spinors and are also given by Brodsky, Pauli and Pinsky~\cite{BrodskyPauliPinsky}. The transverse polarizations are $\epsilon_\pm=\mp(1,\pm i)/\sqrt2$, so $\epsilon_\pm^*\cdot k_\perp$ has an azimuthal phase. For $g\to gg$, the color factors become $T^a\to if^{abc}$ and $C_F\to C_A$. The helicity density matrix in Eq.~\eqref{eq:appb-rhohel} shows which phases survive after the unobserved quark is traced out.

We take the amplitude equation to connect adjacent Fock sectors,
\beq\label{B4}
Q^2 \frac{\partial \psi_n(\{x_i, k_{\perp,i}\}; Q^2)}{\partial Q^2}\;=\; -\,\Sigma_n(Q^{2})\,\psi_n\;+\;
\mathcal R_{n,n-1}(Q^{2})\,\psi_{n-1}\,.
\eeq
Here $n$ labels the Fock sector. The operator $\mathcal R_{n,n-1}$ splits one parton, and $\Sigma_n$ supplies the virtual contribution. 
The leading logarithmic $\Sigma_n$ is half the sum of the differential branching rates of the partons in that sector. The factor one half converts a probability loss into an amplitude loss. The rate $\Sigma_n$ depends only on the parton species for parent momentum fractions $x_p\geq x_{\rm min}/\varepsilon$. The rates are those of Eq.~\eqref{B5001}, cut at the endpoint resolution $\varepsilon$ of the $z$ integrals (Appendix~\ref{AppB}).
Between emissions the amplitude contains the factor $\sqrt\Delta$. The no-emission probability of one parent of a given species is $\Delta(Q_0^2,Q^2)=\exp[-\int_{Q_0^2}^{Q^2}(dq^2/q^2)\int_\varepsilon^{1-\varepsilon} dz\,(\alpha_s/2\pi)\hat P^{(0)}(z)]$.
The parent's total real kernel is $\hat P^{(0)}=\hat P^{(0)}_{qq}$ for a quark and $\hat P^{(0)}=\tfrac12\hat P^{(0)}_{gg}+n_f\hat P^{(0)}_{qg}$ for a gluon.
The longitudinal integral uses the cut range of Appendix~\ref{AppB}. The resolution cutoffs are removed in the DGLAP limit of Section~\ref{SecBDilute}.

The real kernel conserves longitudinal and transverse momentum at each splitting. 
Each unordered daughter pair appears once, replacing a parent $p$ of species $j$ by daughters $b,c$. The daughter $b$ has species $i$ and relative fraction $z$, which fixes the labeling of $\mathcal V_{ij}$. The kernel reads

\beqar\label{B6001}
&\,&
\bigl[\mathcal R_{n,n-1}\,\psi_{n-1}\bigr](\{x_i, k_{\perp,i}\};Q^{2})\,=\,\sum_{\rm splittings}\;
\int_{0}^{1}\!\! dz\int\! d^{2}k_\perp
\int_{0}^{1}\!dx_{p}\int\! d^{2}k_{\perp,p}\;
4\pi\sqrt{x_{p}}\;\mathcal V_{ij}(z, k_\perp) \nn 
\\
&&\times\;\delta\bigl(x_{b}-z\,x_{p}\bigr)\,\delta\bigl(x_{c}-(1-z)x_{p}\bigr)\,
\delta^{2}\bigl(k_{\perp,b}-z\,k_{\perp,p}-k_{\perp}\bigr)\,
\delta^{2}\bigl(k_{\perp,c}-(1-z)k_{\perp,p}+k_{\perp}\bigr) \nn 
\\
&&\times\;\delta\bigl(k_{\perp}^{2}-Q^{2}\bigr)\,k_{\perp}^{2}\;
\psi_{n-1}\bigl(x_{p},k_{\perp,p};\{x_i, k_{\perp,i}\}_{i\neq b,c}\bigr)\,,
\eeqar
with the spectator arguments unchanged.
The factor $4\pi\sqrt{x_p}$ converts the rate vertex of Eq.~\eqref{B2} to the Fock normalization of the measure~\eqref{AA3}. The delta functions give the boundary amplitude $(4\pi/\sqrt{x_p})\,\mathcal V_{ij}\,\psi_{n-1}$. Its square integrated with the measure~\eqref{AA3} gives the real emission probability $(\alpha_s/2\pi)\hat P^{(0)}_{ij}(z)\,dz\,dk_\perp^2/k_\perp^2$ for every parent fraction $x_p$. Each term includes the ratio of the symmetry factors of the two Fock sectors, given in Eq.~\eqref{AppB6001001}. The observed daughter is selected in the density of Eq.~\eqref{AppB600301}.

 Equation~\eqref{B4} is an amplitude ansatz for strongly ordered transverse momenta in light-cone gauge. Its boundary kernel creates orthogonal components at different scales. The prescription $\theta(0)=1/2$ gives the boundary term in Eq.~\eqref{B4001}, so the summed norm is conserved with the matched virtual loss of Eq.~\eqref{B11}.
The squared amplitudes reproduce leading-order DGLAP as the cutoffs on momentum fractions are removed with $x\geq x_{\rm min}/\varepsilon$ maintained (Sec.~\ref{SecBDilute} and Appendix~\ref{AppB}). This reduction holds at leading logarithmic accuracy for strongly ordered transverse momenta in light-cone gauge, dilute occupations, including those of newly resolved modes, and finite mean parton number.
 The solution sums all ordered emission histories, whereas the amplitude for a fixed sequence decays between its emissions. We separately average $|\psi\rangle\langle\psi|$ over unrecorded histories to define the history-averaged state $\hat\rho$. This additional assumption discards their coherences and gives a mixed state. It is not needed for the boundary norm balance. The fixed-mode equation~\eqref{B10} uses the shell operator without the scale delta function and cannot conserve every state's norm, as shown after Eq.~\eqref{B10001}.

The diagonal DGLAP equation leaves scale-dependent phases of the amplitudes undetermined. Our amplitude model admits no additional such phase beyond those of the chosen vertices. The evolved helicity off-diagonal elements of $\hat\rho_1$ can depend on this convention beyond leading logarithmic accuracy. 
Later occupations can depend on this phase convention through coherences unless the phase generator commutes with every occupation operator and its rotation leaves the dissipator invariant (Section~\ref{SecBC}).
 The model also does not resolve the interference of soft emissions from different parton legs. That interference is distinct from the helicity coherence in $\hat\rho_1$ and from the coherence between channels of a single splitting considered below. Appendix~\ref{AppC} gives the first emission amplitudes and their ordered scale integrals.

\subsection{Emission histories and the density operator}\label{SecBC}

Each recorded emission history specifies a pure state after normalization. Strong ordering makes the evolution local in the current scale $t=\ln Q^2$ at leading logarithmic accuracy. This locality does not fix the interference between histories that lead to the same final configuration. Identifying the chosen history ensemble with the full coherent hadron wave function would require their overlaps and a common measure, which we do not supply.

The splitting operator on the parton Fock space is
\beq\label{B9}
\hat{\mathcal R}\,=\,\sum_{a,b,c}V_{ab,c}\;a^{\dagger}_{a}\,a^{\dagger}_{b}\,a_{c}\,,
\eeq
which annihilates the parent mode $c$ and creates the daughter modes $a,b$. The coefficient $V_{ab,c}$ contains the vertex, its momentum constraints and the normalization of the ordered mode sums. For a two-daughter channel $V_{ab,c}=\mathcal V_{ab,c}/S_V$, with $S_V=2$. Only the part of $V$ with the exchange symmetry of the two creation operators contributes. It is symmetric when the operators commute, as for $g\to gg$ and the quark and gluon of $q\to qg$, and antisymmetric when both are fermionic, as for the quark and antiquark of $g\to q\bar q$. The matrix element between normalized Fock states is $\langle ab|\hat{\mathcal R}|c\rangle=S_VV_{ab,c}=\mathcal V_{ab,c}$ for distinct daughter modes. Appendix~\ref{AppB} gives the counting in the ordered sums and its extension to higher multiplicities.

The operator in Eq.~\eqref{B9} acts on a shell at $k_\perp^2=e^t$. Its coefficients depend on $z$, azimuth and species, and it can split any parton in the state. There is no integration over the shell radius and no scale delta function. The latter belongs to the boundary kernel in Eq.~\eqref{B6001}. Thus $\hat{\mathcal R}^\dagger\hat{\mathcal R}$ is a rate operator and contains no squared distribution. We keep the mode and occupation regulators while using these operator products, with finite rates and expectation values. Removing the regulators requires common domains for the unbounded operators, finite rates and convergence of the regulated states and generators that preserves positivity and trace.

Between splittings the unnormalized state obeys
\beq\label{B10002}
Q^{2}\frac{\partial}{\partial Q^{2}}\,|\psi\rangle\,=\,-\,\hat\Sigma\,|\psi\rangle\,,
\eeq
and its norm loss over $dt$ is $\langle\hat\Sigma+\hat\Sigma^\dagger\rangle dt$ to first order for a normalized initial state. The emission probability is $\langle\hat{\mathcal R}^\dagger\hat{\mathcal R}\rangle dt$. Agreement for every state fixes the Hermitian part of $\hat\Sigma$ to $\hat{\mathcal R}^\dagger\hat{\mathcal R}/2$. We take $\hat\Sigma$ Hermitian, with its possible anti-Hermitian part discussed below Eq.~\eqref{B12}, so
\beq\label{B11}
\hat\Sigma\;=\;\frac{1}{2}\,\hat{\mathcal R}^{\dagger}\hat{\mathcal R}\,.
\eeq
This $\hat\Sigma$ is a three-body operator. 
Its expectation value reduces to the rate $\Sigma_n$ of Eq.~\eqref{B4} when the daughter modes are empty.
 Dalibard, Castin and M\o lmer~\cite{DalibardCastinMolmer} and Carmichael~\cite{Carmichael} use this jump and no-jump description in quantum trajectory theory. Nagy and Soper~\cite{NagySoper} fix virtual terms from probability conservation in parton shower density evolution, retaining interference between emissions from different legs, and generalize the shower to any order in $\alpha_s$. The integrated relation between real and virtual contributions is discussed by Ioffe, Fadin and Lipatov~\cite{IoffeFadinLipatov}. Curci, Furmanski and Petronzio~\cite{CFP} obtained the endpoint term of the nonsinglet kernel from the real diagrams through fermion-number conservation at next-to-leading order. Here the probability condition of the jump and no-jump description is applied to the collinear splitting operators on parton Fock space.

The condition weakens to the integrated balance of Eq.~\eqref{B4001} if imposed only on the physical hadron state. The two formulations agree at leading logarithmic accuracy, not necessarily beyond. Pl\"atzer~\cite{PlatzerColour} keeps the real and virtual operators unrelated because a definite projection onto final states can prevent a one-to-one relation between them. Equation~\eqref{B11} concerns the unprojected evolution. The soft-gluon evolution of \'Angeles Mart\'inez et al.~\cite{AngelesMartinez} contains a virtual operator equal to one half the square of its emission operator, with the structure of Eq.~\eqref{B11}.

The history average can now be evolved over one scale step. To first order in $dt$, the no-emission branch contributes $\hat\rho-\{\hat\Sigma,\hat\rho\}dt$, and the emission branch contributes $\hat{\mathcal R}\hat\rho\hat{\mathcal R}^\dagger dt$. Their sum gives
\beq\label{B12}
\frac{d\hat\rho}{dt}\,=\,\hat{\mathcal R}\,\hat\rho\,\hat{\mathcal R}^{\dagger}
\,-\,\frac{1}{2}\Bigl\{\hat{\mathcal R}^{\dagger}\hat{\mathcal R},\,\hat\rho\Bigr\}\,=\,
\hat{\mathcal R}\,\hat\rho\,\hat{\mathcal R}^{\dagger}
\,-\,\Bigl\{\hat\Sigma,\,\hat\rho\Bigr\}\,,        
\qquad t\,=\,\ln Q^{2}\,
\eeq
which has the form of density evolution given by Lindblad~\cite{Lindblad} and Gorini, Kossakowski and Sudarshan~\cite{GKS}. The regulated equation preserves positivity and trace. These finite-regulator statements do not establish the continuum limit described above.

Armesto et al.~\cite{ArmestoCGC} derived Lindblad evolution in rapidity for a Color Glass Condensate density matrix whose retained variable is the rapidity-integrated valence color charge.
 Li and Kovner~\cite{LiKovner} obtained Lindblad evolution of the hard-gluon density matrix in the dense and dilute limits. They also noted that a Markovian evolution need not have this form when its evolution variable labels the modes being integrated out. Gu et al.~\cite{GuRG} used the renormalization scale as the evolution parameter for final-state spin decoherence.
The Lindblad equation of Ref.~\cite{Rey2026} describes leading-soft QED evolution in the logarithm of the photon energy resolution below which photons are traced out, with virtual terms matched to eikonal emission terms.
Its diagonal jump operators dephase a fixed set of hard QED configurations without changing their probabilities or the number of hard particles.
 Chachamis, Hentschinski and Sabio Vera~\cite{ChachamisLindblad} modeled saturation in the Green function of Balitsky, Fadin, Kuraev and Lipatov (BFKL)~\cite{BFKL} through a non-Hermitian, quasi-Lindbladian evolution.
Here the Lindblad form follows from the history average and the matching of emission and no-emission probabilities.

Probability conservation also permits an anti-Hermitian contribution $i\hat H_C$ to $\hat\Sigma$, with $\hat H_C$ Hermitian. It adds the commutator in Eq.~\eqref{B12001} and corresponds to the Lamb shift of open-system evolution described by Breuer and Petruccione~\cite{BreuerPetruccione}, fixed by the dispersive amplitudes. The $i\pi$ part of the one-loop virtual contribution in the shower of Nagy and Soper~\cite{NagySoper} is such a phase term. It enters their full generator and is usually omitted in practice, while probability conservation does not fix it. \'Angeles Mart\'inez et al.~\cite{AngelesMartinez} separate this Coulomb term from the radiative virtual operator, and Forshaw, Holguin and Pl\"atzer~\cite{FHP} retain Coulomb, or Glauber, exchanges in amplitude-level parton branching with soft and collinear emissions. Their unitarity relation gives the virtual contribution, and their collinear limit recovers DGLAP evolution.

A mode-diagonal phase generator makes no direct contribution to the instantaneous occupation derivatives. It acts through coherences between configurations reached by different histories, which the leading logarithmic history average discards. We set $\hat H_C=0$. 
A phase generator commuting with every occupation operator leaves occupations unchanged under its rotation. A dissipator invariant under that rotation permits setting $\hat H_C=0$ without changing their evolution.
 Otherwise, this choice is an assumption of the model.

\subsection{The dilute DGLAP limit}\label{SecBDilute}

The occupations from Eq.~\eqref{B12} obey leading-order equations for observed fractions $x\geq x_{\rm min}/\varepsilon$. Every gain and loss integral is cut to $\varepsilon<z<1-\varepsilon$. The endpoint logarithms cancel. The cut occupation equations give Eq.~\eqref{AA1} as the endpoint cutoff vanishes. The limit keeps $x\geq x_{\rm min}/\varepsilon$ and a dilute new shell, $[\gamma_G+2\Sigma_g(\varepsilon)]\,\bar n\ll1$ (Appendix~\ref{AppB}). This reduction holds at leading logarithmic accuracy for strongly ordered transverse momenta in light-cone gauge. Strong ordering places both daughters in empty modes of the new shell at each splitting in the dilute-shell regime described below. The six-operator expectation value in Section~\ref{SecBOccupations} then reduces to the parent occupation. Small mean occupations alone do not suffice because parent-daughter correlations can contribute at that same order.

We require dilute occupations $n_\lambda\ll1$ and a dilute new shell to keep its Bose enhancement small (Appendix~\ref{AppB}). The occupation $n_{\rm shell}$ counts partons per newly resolved mode. This occupation must be much smaller than one. Section~\ref{SecResolution} gives the bound on mean occupation for the gluon example.
 The virtual term removes the observed parent at its full branching rate. Emissions on unobserved spectators cancel their virtual losses after the Fock sectors are summed. The two daughter contributions give the real convolution with $\hat P^{(0)}_{ij}$, and the endpoint cancellation supplies the plus prescription and the $\delta(1-z)$ terms. 
The cutoff guarantees finite mean multiplicity away from $x=0$. Appendix~\ref{AppB} gives the boundary calculation and the limiting coefficients $3C_F/2$ and $(11C_A-4n_fT_F)/6$.

The spectator cancellation is the real-virtual cancellation on unobserved partons of the jet calculus of Konishi, Ukawa and Veneziano~\cite{KUV}. The reduction is similar to their derivation and to that of Ioffe, Fadin and Lipatov~\cite{IoffeFadinLipatov}.

This reduction uses the real kernels as input through Eq.~\eqref{B2}. It tests their normalization against the virtual rates, the quark-number sum rule and the momentum sum rules. Beyond leading order, the same-final-state interference, subtractions and factorization scheme must be specified before an inclusive kernel can be obtained.

\subsection{Occupations and the thermal Gaussian approximation}\label{SecBOccupations}

The evolution of a one-body occupation depends on higher moments of the state. For $N_\lambda=a^\dagger_\lambda a_\lambda$ and $n_\lambda=\Tr[\hat\rho N_\lambda]$, Eq.~\eqref{B12} gives a six-operator expectation value containing the parent and both daughters. It is the first equation of the hierarchy of reduced density operators described in Appendix~\ref{AppD}. The mode labels here refer to a fixed momentum basis, and $n_\lambda$ is the dimensionless occupation of a specified cell. The longitudinal cell measure used in the entropy-sign test is stated in Appendix~\ref{AppB}. The diagonal occupations in this basis coincide with the natural-orbital eigenvalues of Section~\ref{SecA} only when the one-body matrix is diagonal in that basis.

The thermal Gaussian approximation evaluates the occupation moments in the state of Eq.~\eqref{AB18}, with coherence between distinct mode sets neglected. This neglect is separate from block diagonality in $x$. The connected normal cumulants $\hat C_m$ are the parts of the reduced operators $\hat\rho_m$ left after products of lower-rank correlations are subtracted (Appendix~\ref{AppD}). The thermal Gaussian state sets these cumulants above rank one to zero,
\beq\label{B15001}
\hat C_{m}\,=\,0\qquad\hbox{for all}\ m\geq2\,,
\eeq
but their vanishing does not imply that state (Section~\ref{SecAMode}). Equation~\eqref{B16} uses only these number-conserving cumulants to give the distinct-mode expectation value
\beq\label{B16}
\Bigl\langle a^{\dagger}_{c}a_{b}a_{a}\,a^{\dagger}_{a}a^{\dagger}_{b}a_{c}\Bigr\rangle
\;\longrightarrow\;n_{c}\,\bigl(1+n_{a}\bigr)\bigl(1+n_{b}\bigr)\,.
\eeq
The Bose factors come from $a_a a_a^\dagger=1+a_a^\dagger a_a$. They remain after the connected cumulants are set to zero. For coincident daughters, the thermal contraction is $\langle a_a^2a_a^{\dagger2}\rangle=2(1+n_a)^2$. If all three modes coincide, the value is $2n(n+1)(3n+1)$. The momentum delta functions of a splitting at nonzero $Q$ exclude those coincidences for distinct continuum modes. 
The residual coincidence weight for resolved cells is $O(1/N_\perp)$ if the daughters are spread over the $N_\perp$ transverse modes.
 The one-mode comparison is given in Appendix~\ref{AppA}.

The resulting bosonic gain and loss equation is
\beq\label{B17}
\frac{dn_{\lambda}}{dt}\,=\,\frac{2}{S_{V}}\sum_{b,c}\bigl|\mathcal V_{\lambda b,c}\bigr|^{2}\,
n_{c}\bigl(1+n_{\lambda}\bigr)\bigl(1+n_{b}\bigr)
\;-\;\frac{1}{S_{V}}\sum_{a,b}\bigl|\mathcal V_{ab,\lambda}\bigr|^{2}\,
n_{\lambda}\bigl(1+n_{a}\bigr)\bigl(1+n_{b}\bigr)\,,
\eeq
where the factor two in the gain counts the two daughter slots in which the observed mode can appear. The ordered sums and the normalization of $V$ leave the explicit $1/S_V$. Quark modes have $1-n$ in place of $1+n$, with the statistics assigned separately to every species. Thus $q\to qg$ has the factor $n_c(1-n_q)(1+n_g)$.

The Gaussian approximation does not follow from the vertex truncation because even a single $1\to2$ vertex generates connected correlations. 
We apply the approximation to color-averaged occupations, estimating its error only for the initial evolution of one bosonic channel $c\to a+b$ with distinct modes, a thermal parent of mean occupation $n$ and empty daughters.
The full state's splitting rate initially equals its factorized form $|\mathcal V_{ab,c}|^2n_c(1+n_a)(1+n_b)$, with initial derivatives in $t$ of $|\mathcal V_{ab,c}|^4(6n^2-n)$ from the Lindblad equation~\eqref{B12} and $|\mathcal V_{ab,c}|^4(2n^2-n)$ from the closed occupation equation~\eqref{B17}.
The derivatives agree to first order in $n$, before Bose factors contribute, but differ by $4|\mathcal V_{ab,c}|^4n^2$ because the approximation discards correlations created among the three modes by splitting.
Their relative difference is $4n$ for small $n$, and their signs are opposite for $1/6<n<1/2$, so dilute occupations are necessary but do not suffice.
 The correlations imposed by color and momentum constraints are discussed in Section~\ref{SecAMode}.

Duan, Kovner and Lublinsky~\cite{DuanBORG} observed that DGLAP produces the partons with the largest transverse momentum only in the last splitting. Previously created partons therefore do not populate the new shell at leading logarithmic accuracy. 
The source $W$ is the occupation a newly resolved shell mode acquires without Bose enhancement, identified with $n_{\rm shell}$ of Section~\ref{SecResolution}. Appendix~\ref{AppB} gives the fixed-source prescription, its Bose enhancement and the unchanged leading logarithmic kernel. The dilute expansion fails at the occupancy crossover for $\varepsilon\lesssim10^{-2}$.

Pauli and Bose factors for both daughters also appear in the leading-order equations of Ref.~\cite{ManganoMieleMigliore}, introduced by analogy with the Boltzmann equation and not derived from a state.
The Bose factor $1+n_\lambda$ has the statistical origin of the stimulated emission in Ref.~\cite{DuanBORG}. 
Our Bose factor acts only within the newly resolved shell, whereas the Born-Oppenheimer cascade ordered in light-cone frequency contains partons of high transverse momentum at each step if their longitudinal momentum is high enough.
Duan, Kovner and Lublinsky~\cite{DuanBORG} identify their stimulated emission as specific to their cascade. They caution that their dilute analysis may require modification at very small $x$.

\subsection{Entropy flow}\label{SecBEntropy}

The mode entropy follows the populations in the thermal Gaussian approximation with mode coherence neglected. Equations~\eqref{B17001} through~\eqref{B17003} therefore evolve the entropy of the dephased momentum and helicity occupations. This value is an upper bound on the mode-entropy functional of the coherent one-body matrix in Section~\ref{SecAMode}. Its derivative in the dephased basis is
\beq\label{B17001}
\frac{dS_{\rm mode}}{dt}\,=\,\sum_{\lambda}\epsilon_{\lambda}\,\frac{dn_{\lambda}}{dt}\,,\qquad
\epsilon_{\lambda}\,=\,\ln\Le\frac{1+n_{\lambda}}{n_{\lambda}}\Ra\,,
\eeq
where the weights $\epsilon_\lambda$ are the modular energies of Eq.~\eqref{AB19}. One splitting $c\to a+b$ increases the daughter occupations and decreases the parent occupation, so Eq.~\eqref{B17} gives
\beq\label{B17002}
\frac{dS_{\rm mode}}{dt}\,=\,\sum_{\rm vertices}R_{f}\,
\bigl(\epsilon_{a}+\epsilon_{b}-\epsilon_{c}\bigr)\,=\,
\sum_{\rm vertices}R_{f}\,\ln\frac{R_{f}}{R_{b}}\,,
\eeq
where
\beq\label{B17003}
R_{f}\,=\,\bigl|\mathcal V_{ab,c}\bigr|^{2}\,n_{c}\bigl(1+n_{a}\bigr)\bigl(1+n_{b}\bigr)\,,\qquad
R_{b}\,=\,\bigl|\mathcal V_{ab,c}\bigr|^{2}\,n_{a}n_{b}\bigl(1+n_{c}\bigr)\,,
\eeq
where $R_f$ is the splitting rate, and $R_b$ is the inverse rate that would accompany it in detailed balance.
 The generator used here contains splittings only. The inverse channel is absent by choice of generator, independently of the Gaussian approximation.
The entropy rates hold for $0<n_\lambda$. An initially empty mode has a logarithmically large entropy rate as it fills, although its entropy stays continuous.

The contribution of one channel to Eq.~\eqref{B17002} is nonnegative when $n_c(1+n_a+n_b)\geq n_a n_b$. The dilute bracket is $\ln[n_c/(n_an_b)]$ to leading order, so dilution alone does not fix its sign. The power-law test in Appendix~\ref{AppB} states the longitudinal cell measure and the common transverse prefactor needed to apply this condition. If both splitting and merging were included with the same statistical factors, the entropy contribution would be $(R_f-R_b)\ln(R_f/R_b)\geq0$. The effective kinetic theory of Arnold, Moore and Yaffe~\cite{AMY} includes both directions of the collinear process with Bose and Pauli factors.

For quarks, the weight is $\epsilon_\lambda=\ln[(1-n_\lambda)/n_\lambda]$ for $n_\lambda<1$, and each fermionic final-state factor is $1-n$.
Equation~\eqref{B28} still bounds the von Neumann entropy of a state outside the thermal Gaussian family if the maximizing state exists and $S_{\rm mode}$ is finite. The occupation flow then determines the bound's change, not the entropy production of the full state. Appendix~\ref{AppD} gives the covariance needed for a paired Gaussian state.

\subsection{Several splitting channels and momentum conservation}\label{SecBD}

Higher-order collinear amplitudes can produce more than two daughters. The triple-collinear splitting functions of Catani and Grazzini~\cite{CataniGrazzini} give the first such case at the level of squared matrix elements. Those functions do not determine the splitting amplitudes or their phases. We denote by $\hat{\mathcal R}_k$ a vertex that replaces one parent by $k+1$ daughters,
\beq\label{B19}
\hat{\mathcal R}\,=\,\sum_{k\geq1}\hat{\mathcal R}_{k}\,,\qquad
\hat{\mathcal R}_{k}\,=\,\sum V^{(k)}_{a_{1}\ldots a_{k+1},\,c}\;
a^{\dagger}_{a_{1}}\cdots a^{\dagger}_{a_{k+1}}\,a_{c}\,,
\eeq
where the splitting vertices remain inputs. The probability condition fixes their virtual partners after a prescription for coherence between channels has been chosen.

H\"oche and Prestel~\cite{HochePrestel} included subtracted $1\to3$ splitting kernels in a parton shower, first for the flavor-changing channels. The shower's unitarity supplies the approximate virtual and unresolved real corrections. These branchings only raise the parton multiplicity.

A single jump operator $\hat{\mathcal R}=\sum_k\hat{\mathcal R}_k$ adds the channels coherently. A separate jump operator for each distinguishable channel gives a sum of dissipative terms. The corresponding no-emission operators are
\beq\label{B19001}
\hat\Sigma\,=\,\frac{1}{2}\sum_{k,k'\geq1}\hat{\mathcal R}^{\dagger}_{k}\hat{\mathcal R}_{k'}\,,
\qquad\hbox{against}\qquad
\hat\Sigma\,=\,\frac{1}{2}\sum_{k\geq1}\hat{\mathcal R}^{\dagger}_{k}\hat{\mathcal R}_{k}\,,
\eeq
and both choices preserve trace and positivity for the regulated evolution. Norm conservation does not select between them. They give the same norm decrement, total rate and population derivatives at the instant the state has definite parton number, as Eq.~\eqref{B19002} shows. Their evolved density matrices differ because the coherent generator also contains $\hat{\mathcal R}_k\hat\rho\hat{\mathcal R}_{k'}^\dagger$ and $-\tfrac12\{\hat{\mathcal R}_{k'}^\dagger\hat{\mathcal R}_k,\hat\rho\}$ for $k\neq k'$. These terms connect different Fock sectors and can affect subsequent populations.

Separately resolved channels give
\beq\label{B19003}
\frac{d\langle\hat N\rangle}{dt}\,=\,\sum_{k\geq1}k\,
\bigl\langle\hat{\mathcal R}^{\dagger}_{k}\hat{\mathcal R}_{k}\bigr\rangle\,\geq\,0\,,
\eeq
which holds for every state, so the multiplicity cannot decrease for separately resolved channels. Their generator cannot represent the fusion of two partons into one. Coherent addition gives
\beq\label{B19004}
\frac{d\langle\hat N\rangle}{dt}\,=\,\sum_{k,k'\geq1}\frac{k+k'}{2}\,
\bigl\langle\hat{\mathcal R}^{\dagger}_{k}\hat{\mathcal R}_{k'}\bigr\rangle\,,
\eeq
whose cross terms can have either sign. 
The virtual cross terms for coherent addition change the parton number by $k'-k$, raising and lowering it, as Eqs.~\eqref{B20} and~\eqref{B21} show. The lowering terms are not a recombination of densities, and their coefficients are fixed by the supplied splitting amplitudes.

Such a change of Fock sector does not derive the twist-four recombination mechanism of Gribov, Levin and Ryskin~\cite{GLR} or Mueller and Qiu~\cite{MuellerQiu}. An independent merging jump would have its own amplitude, and probability conservation would fix only its virtual partner. No recombination kernel or twist-four operator matching is obtained here. Products with different $k$ also do not give the inclusive next-to-leading-order real and virtual interference, which pairs amplitudes with the same final state. The loop counting and the required subtraction of iterated lower-order splittings are stated in Appendix~\ref{AppD}. Kutak and Prasza{\l}owicz~\cite{KutakPraszalowicz} study a one-dimensional cascade with splitting and recombination.

The operator $\hat P^+=\sum_\lambda x_\lambda N_\lambda$ measures longitudinal momentum in units of the hadron momentum. Its mean is conserved for every retained set of momentum-conserving vertices with their matched virtual terms. The sum in the conservation law
\beq\label{B24005}
\sum_{\lambda}x_{\lambda}\,\frac{dn_{\lambda}}{dt}\,=\,0\,,
\eeq
runs over the retained modes $x\geq x_{\rm min}$ of Appendix~\ref{AppB}. Its vanishing follows because the adjoint of Eq.~\eqref{B12} gives
\beq\label{eq:appc-4}
\frac{d\langle\hat P^{+}\rangle}{dt}\,=\,\frac{1}{2}\sum_{k,k'\geq1}
\Bigl[\,\Bigl\langle\hat{\mathcal R}^{\dagger}_{k'}\bigl[\hat P^{+},\hat{\mathcal R}_{k}\bigr]\Bigr\rangle
\,+\,\Bigl\langle\bigl[\hat{\mathcal R}^{\dagger}_{k'},\hat P^{+}\bigr]\hat{\mathcal R}_{k}\Bigr\rangle\Bigr]\,,
\eeq
and the momentum delta functions set $[\hat P^+,\hat{\mathcal R}_k]=0$ for each $k$, so the right-hand side vanishes for any state, including states with connected or anomalous correlations. An inconsistent truncation can break this result by keeping a gain without its matched loss or by evaluating the paired terms in different states.

The corresponding constraint on a merging vertex is stated after Eq.~\eqref{B19005}. Momentum conservation is not automatic in nonlinear evolution. Duwent\"aster, Guzey, Helenius and Paukkunen~\cite{NonlinearFit} note that the recombination equations of Gribov, Levin and Ryskin and of Mueller and Qiu violate it. Their global fit uses the equations of Zhu and Ruan~\cite{ZRS}, which conserve momentum through the combined shadowing and antishadowing terms.

The same argument applies to baryon number and electric charge when each vertex conserves them. The thermal Gaussian approximation, with coherence between distinct mode sets neglected, gives the channel-diagonal occupation equation~\eqref{B24}. More general states require the interference moments in Eq.~\eqref{B24002}. Odd products vanish for a Gaussian state with vanishing one-point functions because that state preserves parton-number parity. A non-Gaussian state may violate this condition, as discussed in Appendix~\ref{AppD}.

The relative sizes of successive channel-diagonal terms can be estimated by power counting with vertex coefficients of order one. Each extra daughter costs one power of $\alpha_s$ and contributes one Bose factor. Relative to the $k=1$ term, this gives
\beq\label{B25}
\Bigl[\,\alpha_{s}\,\bigl(1+n_{\rm shell}\bigr)\Bigr]^{k-1}\;
\eeq
where Section~\ref{SecResolution} relates the shell occupation to the mean occupation below $Q$. The weight in Eq.~\eqref{B25} uses the mode count, gluon density and endpoint resolution $\varepsilon$ specified there.

\section{Resolution and scaling}\label{SecResolution}

\subsection{Finite longitudinal cells}

The entropy assigned to resolved partons depends on which modes are distinguished. Separating longitudinal cells, removing coherence between them, and merging their occupations are different operations. We define each operation at finite resolution and assume finite traces and finite entropies. The function $g=g_B$ is the gluon entropy function of \eq{AB21002}.

At fixed $Q^2$, the longitudinal fraction $x$ labels modes of one hadron. The projector onto an $x$ cell selects a subsystem of its Fock space. 
The projector $P_i$ selects the resolved modes of cell $i$. Block diagonality gives
\beq\label{AC2}
S_{\rm mode}\bigl[\hat\rho_1\bigr]\,=\,\sum_{i}s_{\rm mode}(x_i,Q^{2})\,,\qquad
s_{\rm mode}(x_i,Q^{2})\,=\,\Tr\,g\bigl(P_i\hat\rho_1P_i\bigr)\,.
\eeq
In \eq{AC2} and the cell formulas below, $x_i$ labels the specified longitudinal cells and the sums run over these cells. 
The quantity $s_{\rm mode}(x_i,Q^2)$ is a cell entropy, not an entropy density per unit $x$.
 The trace in \eq{AC2} runs over the eigenvalues of each cell block, not its diagonal elements. These eigenvalues are dimensionless occupations. A smooth density per unit $x$ cannot be inserted into $g$ in their place. Block diagonality holds in the translation-invariant packet limit specified in Appendix~\ref{AppAFock}, while \eq{AC4} measures the residual coherence at finite width.

Each longitudinal block still contains the transverse and internal labels. For the diagonal cell estimates we retain one longitudinal mode per cell. The transverse mode count then gives the following form if the block is diagonal in the remaining labels as well,
\beq\label{AC2002}
s_{\rm mode}(x_i,Q^{2})\,=\,\sum_{\rm f,c,s}S_{\perp}\int\frac{d^{2}k_{\perp}}{(2\pi)^{2}}\;
g\bigl(n_i(k_{\perp},{\rm f},{\rm c},{\rm s};Q^{2})\bigr)\,,
\eeq
where $S_\perp$ is the transverse area. The diagonal cell estimate gives $n_i(\mu)=(4\pi S_\perp)^{-1}\int_{{\rm cell}\,i}dx\,n(x,\mu)$, with the kernel $n(x,\mu)$ of \eq{AB16} and $\mu=(k_\perp,{\rm f,c,s})$. The argument of $g$ is this cell occupation. There is one transverse mode per area $(2\pi)^2/S_{\perp}$ in transverse momentum space. The measure agrees with the parton number in \eq{AB16}. Taking the diagonal of a block can increase its mode entropy, so \eq{AC2002} is an upper bound if transverse or internal coherence remains.

Removing coherence between longitudinal cells also increases the mode entropy. Concavity of $g$ gives
\beq\label{AC3}
\sum_{i}s_{\rm mode}(x_i,Q^{2})\;\geq\;S_{\rm mode}\bigl[\hat\rho_1\bigr]\;;
\eeq
the difference of its two sides defines
\beq\label{AC4}
I_{x}(Q^{2})\,\equiv\,\sum_{i}s_{\rm mode}(x_i,Q^{2})\,-\,S_{\rm mode}\bigl[\hat\rho_1\bigr]\,\geq\,0\,,
\eeq
with $I_x$ built from $g$ of the unnormalized occupation matrix. The coherence $C_{\rm rel}$ in \eq{AB17002} uses the entropy of a normalized one-parton matrix and is a different functional. The quantity $I_x$ is the total correlation between the cell Fock spaces for a number-conserving thermal Gaussian state, or their mutual information for two cells. For a general state it measures the change of $S_{\rm mode}$ on removing the cell coherence.

Merging longitudinal cells sums their occupation matrices in the remaining transverse and internal space,
\beq\label{AC6}
\hat\rho_1^{\perp}(k_{\perp},{\rm fcs};k_{\perp}',{\rm fcs}')\,=\,\sum_{i}\hat\rho_1(x_i,k_{\perp},{\rm fcs};\,x_i,k_{\perp}',{\rm fcs}')\,,
\qquad \Tr\hat\rho_1^{\perp}\,=\,\langle N\rangle\,,
\eeq
and assigns the entropy
\beq\label{AC7}
S^{\perp}(Q^{2})\,=\,\Tr\,g\bigl(\hat\rho_1^{\perp}\bigr)\,.
\eeq
The entropy $S^\perp$ is the thermal, maximum-entropy assignment for the merged one-body matrix. It is not the entropy of the total parton count.
 The summed matrix acts on fewer modes. Subadditivity of $g$ gives
\beq\label{AC9}
S^{\perp}(Q^{2})\;\leq\;\sum_{i}s_{\rm mode}(x_i,Q^{2})\;
\eeq
and the difference has the following form if the cell blocks are diagonal in a common transverse and internal basis,
\beq\label{AC10}
\Delta_{x}\,=\,\sum_{\mu}\Bigl[\sum_{j=1}^{K}g\bigl(n_{j}(\mu)\bigr)\,-\,g\Bigl(\sum_{j=1}^{K}n_{j}(\mu)\Bigr)\Bigr]\,\geq\,0\,,\qquad
\mu\,=\,(k_{\perp},{\rm f,c,s})\,,
\eeq
where $n_j(\mu)$ is the dimensionless occupation of longitudinal cell $j$ in mode $\mu$. The gap $\Delta_x$ is the entropy gained by separating partons that differ only in $x$. It depends on the binning and can be positive even for uncorrelated cells. 
Merging assigns one thermal mode to the summed occupation of $K$ modes, so $\Delta_x$ differs from the pinching difference in \eq{AC4}.
 The gap $\Delta_x$ bounds the total correlation of the cells only for a number-conserving thermal Gaussian state whose occupation matrix separates into common transverse or internal sectors, with coherence between cells confined to each sector. The transverse cell volume cancels between the resolved and merged entropies in the dilute regime at fixed longitudinal binning. 
The Bose factors break this cancellation at finite occupation.

One transverse mode containing $M$ partons equally distributed over $K$ cells gives $\Delta_x=K\,g(M/K)-g(M)$. The gap vanishes at $K=1$ and increases on subdivision. Its dilute form, \eq{AC10001} in Appendix~\ref{AppA}, grows as $M\ln K$ when $M/K\ll1$. Figure~\ref{fig:resolution}(a) extends to $K=10^3$. For $M=100$, the dilute expression is 29 percent low at $K=100$ and 1.5 percent low at $K=1000$. Finite resolution is an assumption. A discrete natural-orbital spectrum supplies neither a smallest $x$ cell nor saturation under further subdivision. The coordinate-space entropy studied by Dumitru, Kovner and Skokov~\cite{DKSposition} also depends on the resolution.

\subsection{Rapidity cells}

 In small-$x$ evolution, the rapidity $Y=\ln(1/x)$ is also an evolution parameter beside $t=\ln Q^2$, and changing $Y$ changes the state on the same Fock space. A partial trace requires a tensor factorization in which the traced variable labels a factor. The evolution parameter $Y$ labels no such factor, so integrating along the evolution does not take a partial trace. A sum of cell entropies requires a model in which partons carry rapidity labels, as in the dipole cascade of Mueller~\cite{MuellerDipole} used by Hentschinski, Kutak, P{\l}aczek and Rohrmoser~\cite{HKPRcascade}, and in which the cells are specified as subsystems.

We choose unit rapidity cells. Kharzeev and Levin~\cite{KharzeevLevin} found the mean cascade multiplicity $e^{\omega Y'}$, with $\omega$ the growth rate in \eq{AB23}. Its increase over one unit gives the occupation of the cell beginning at $Y'$,
\beq\label{AC21001}
n_{\rm cell}\,\simeq\,(e^{\omega}\,-\,1)\,e^{\omega Y'}\,
\eeq
where $n_j$ denotes the occupation of the cell beginning at $Y'=j$. For integer $Y$, their sum is $e^{\omega Y}-1$.
Adelaja, Prygarin and Yusuf~\cite{AdelajaPrygarinYusuf} showed that the cascade at constant $\omega$, started from one parton, gives geometric emitted counts of means $n_j$ in individual cells. Their joint law is negative multinomial. The counts are independent Poisson variables at fixed common cascade activity, which has a unit-mean exponential distribution and sets their means proportionally. The normalized second factorial cumulants equal one within a cell and between any two cells.
 Adding their isolated entropies gives the large-$Y$ behavior
\beq\label{AC22}
\sum_{j=0}^{Y-1}g_{B}(n_{j})\,=\,\frac{1}{2}\,\omega\,Y^{2}\,+\,O(Y)\,,
\eeq
while the merged entropy is $g_B(e^{\omega Y}-1)$, which approaches $\omega Y+1$. It equals the geometric multiplicity entropy as a scalar function. Hentschinski, Kharzeev, Kutak and Tu~\cite{HentschinskiEE} discuss this linear growth of the parton-number entropy with rapidity, with the intercept as slope. Identifying it with a joint cell entropy requires a single occupied collective mode. The difference between the isolated-cell sum and this merged assignment is
\beq\label{AC23}
\Delta_{Y}\,=\,\frac{1}{2}\,\omega\,Y^{2}\,+\,O(Y)\,,
\eeq
where the terms proportional to $Y$ depend on the discrete sum. They cannot be obtained by integrating the single-cell asymptote \eq{AC21002} in Appendix~\ref{AppA}. The quadratic coefficient scales with the inverse cell width, so the growth of $\Delta_Y$ depends on the chosen rapidity resolution, as $\Delta_x$ depends on the longitudinal binning.

The resolved functional of the scalar occupation matrix on these cells satisfies
\beq\label{AC23001}
g_{B}(\sum_{i}n_{i})\,\leq\,S_{\rm joint}\,\leq\,\sum_{i}g_{B}(n_{i})\,;\qquad\,S_{\rm joint}=\Tr g(\hat\rho_{1})\,,
\eeq
where the right inequality is the pinching inequality \eq{AC3}, and the left follows from subadditivity of $g$ at fixed trace $\sum_i n_i$. The joint functional is the entropy of the cell state if that state is number-conserving thermal Gaussian. Its longitudinal total correlation then obeys
\beq\label{AC23002}
I_{Y}=\sum_{i}g_{B}(n_{i})-S_{\rm joint}\leq\Delta_{Y}
\eeq
with equality if and only if the occupation matrix has rank at most one. One collective mode can extend across several cells. Independent thermal cells instead have $I_Y=0$, although their merging gap can be nonzero.

The entanglement entropy $S_e$ of the color-charge state discussed in Section~\ref{SecBC} grows at $dS_e/dY=\gamma$ in the dilute Gaussian approximation of Armesto et al.~\cite{ArmestoCGC}, and at half that rate in their saturated approximation. Here $\gamma$ is the leading eigenvalue of the evolution of Balitsky, Fadin, Kuraev and Lipatov~\cite{BFKL}. We compare only the rates, since the density matrices differ. Their separation of entropy growth into growth of the subspace on which the matrix is nonzero and decoherence on that subspace can depend on the basis~\cite{ArmestoCGC}. At leading order $\gamma=\omega=4\ln2\,\bar\alpha_s$, with $\bar\alpha_s=\alpha_sN_c/\pi$.

Prygarin, Sandeu and Yusuf~\cite{PrygarinSandeuYusuf} found the rapidity growth rate $4\ln2\,\bar\alpha_s$ by integrating the conjugated BFKL Green function in the forward limit, at leading order and fixed coupling. The integral of $\tfrac12kk'G(k,k';Y)$ runs over $\delta=\ln(k'^2/k^2)$ and the relative azimuth, and its growth rate is the value used here for $\omega$.

\begin{figure}[htbp]
\centering
\includegraphics[width=\textwidth]{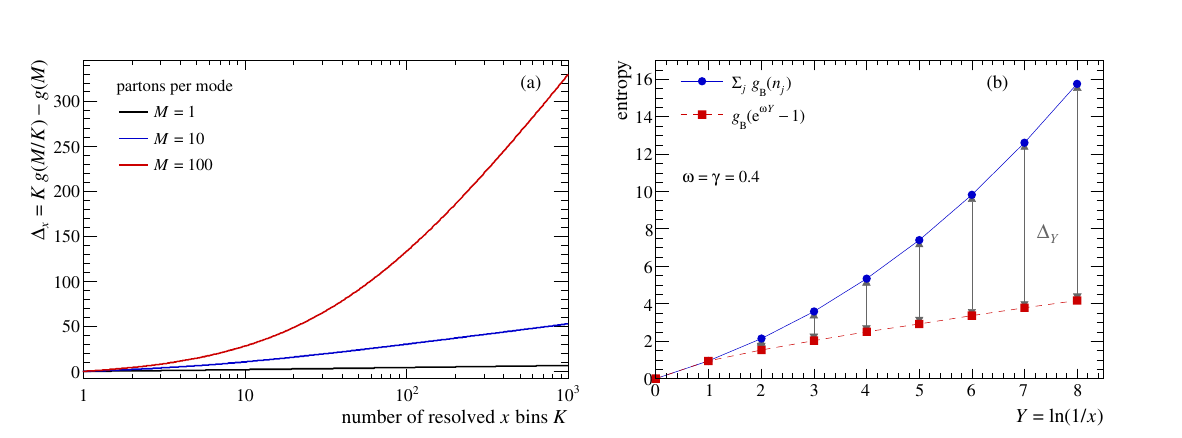}
\caption{Different longitudinal cell models. (a) Gap $\Delta_x=K\,g(M/K)-g(M)$, \eq{AC10}, versus resolved $x$-cell count $K\leq10^3$, for $M=1,10,100$ partons in one transverse mode (black, blue, red). The dilute form \eq{AC10001} requires $M/K\ll1$. (b) Isolated-cell sum $\sum_jg_B(n_j)$ (blue circles) and merged $g_B(e^{\omega Y}-1)$ (red squares) at integer $Y$, for unit cells of \eq{AC21001} and illustrative rate $\omega=\gamma=0.4$, the leading-order $4\ln2\,\bar\alpha_s$ at $\bar\alpha_s\simeq0.14$. Arrows give $\Delta_Y$ on the cell grid. Joining lines guide the eye.}
\label{fig:resolution}
\end{figure}

Figure~\ref{fig:resolution} compares the two finite-cell examples, with panel (b) evaluated at the illustrative rate $\omega=\gamma=0.4$, corresponding to $\bar\alpha_s\simeq0.14$.
 Direct evaluation of the unit-cell sum gives $\Delta_Y=0.60$, $1.55$, $4.47$, and $11.6$ for $Y=2$, $3$, $5$, and $8$. The isolated-cell sum exceeds the merged entropy by factors of $1.8$, $2.5$, and $3.8$ at $Y=3$, $5$, and $8$. These are differences between entropy assignments with the same occupations. They equal the overcount relative to the joint Gaussian entropy only in the rank-one case.  

\subsection{The occupancy crossover}

We now count one longitudinal cell per unit rapidity. The transverse mode count below $Q$ is
\beq\label{AC19}
N_{\perp}\,\simeq\,\frac{S_{\perp}Q^{2}}{4\pi}\times 2(N_{c}^{2}-1)\,,
\eeq
and the average occupation is
\beq\label{AC190001}
\bar n\,=\,xG(x,Q^{2})/N_{\perp}\,.
\eeq
Each unit-rapidity cell is taken as one longitudinal mode, so $s_{\rm mode}$ is the entropy of one such cell and $\bar n$ is its mean occupation per transverse mode. The gluon number in that cell is $\int_{Y_i}^{Y_i+1}dy\,x(y)G(x(y),Q^2)$. The symbol $xG$ in these estimates denotes the corresponding cell-averaged density, equal to the cell number for unit width. We define the saturation scale by
\beq\label{AC19002}
Q_{s}^{2}\,=\,\frac{4\pi^{2}\alpha_{s}N_{c}}{N_{c}^{2}-1}\,\frac{xG(x,Q_{s}^{2})}{S_{\perp}}\,,
\eeq
whose coefficient is four times that of the implicit definition in Eq.~(3.133) of Kovchegov and Levin~\cite{KovchegovLevin}. The two squared saturation scales differ by this factor if the gluon density has the same value at both, as for a density independent of $Q^2$.
 This convention fixes the coefficient in
\beq\label{AC19001}
\bar n(Q^{2})\,=\,\frac{c}{\alpha_{s}}\,\frac{Q_{s}^{2}(x)}{Q^{2}}\,
\frac{xG(x,Q^{2})}{xG(x,Q_{s}^{2})}\,,
\eeq
to $c=1/(2\pi N_c)\simeq0.05$. The occupancy crossover $Q_*$ is defined by $\bar n(Q_*^2)=1$. The fixed-density result is Eq.~\eqref{AC19001001}, obtained when the gluon ratio in \eq{AC19001} is one. The average occupation at $Q_s$ is $c/\alpha_s$, or $0.27$ at $\alpha_s=0.2$.

The scale derivative of the gluon number defines
\beq\label{AC19003}
n_{\rm diff}(Q)\,=\,\frac{\partial (xG)/\partial\ln Q^{2}}{\partial N_{\perp}/\partial\ln Q^{2}}\,=\,
\gamma_{G}\,\bar n(Q)\,,\qquad \gamma_{G}\,=\,\frac{\partial\ln xG}{\partial\ln Q^{2}}\,,
\eeq
and this ratio equals the average occupation only at $\gamma_G=1$. It obeys $n_{\rm diff}<\bar n$ for $\gamma_G<1$, as in the power-law example below. The shell occupation exceeds $n_{\rm diff}$ because each splitting places both daughters in the new shell and removes its parent from an existing mode. The parent loss occurs at the rate $2\Sigma_g$ of Eq.~\eqref{B5001}, with endpoint resolution $\varepsilon$ restricting the integrals to $\varepsilon<z<1-\varepsilon$ (Appendix~\ref{AppB}). The real emissions therefore give $n_{\rm shell}=n_{\rm diff}+2\Sigma_g\bar n=[\gamma_G+2\Sigma_g(\varepsilon)]\,\bar n$, which grows as $\ln(1/\varepsilon)$. The dilute-shell condition $n_{\rm shell}\ll1$ is therefore equivalent to $\bar n\ll1/[\gamma_G+2\Sigma_g(\varepsilon)]$.
 The convention \eq{AC19002} relates $xG$ to $Q_s^2$ along the saturation line, which constrains the total derivative in $x$ and not $\gamma_G$. The slope $\gamma_G$ describes evolution in $Q^2$ and differs from the rapidity growth rate $\gamma$ used above. A power law $xG\propto(Q^2)^{\gamma_G}$ at fixed coupling and constant $\gamma_G\ne1$ gives the crossover $Q_*^2=(c/\alpha_s)^{1/(1-\gamma_G)}Q_s^2$.

\begin{figure}[htbp]
\centering
\includegraphics[width=0.62\textwidth]{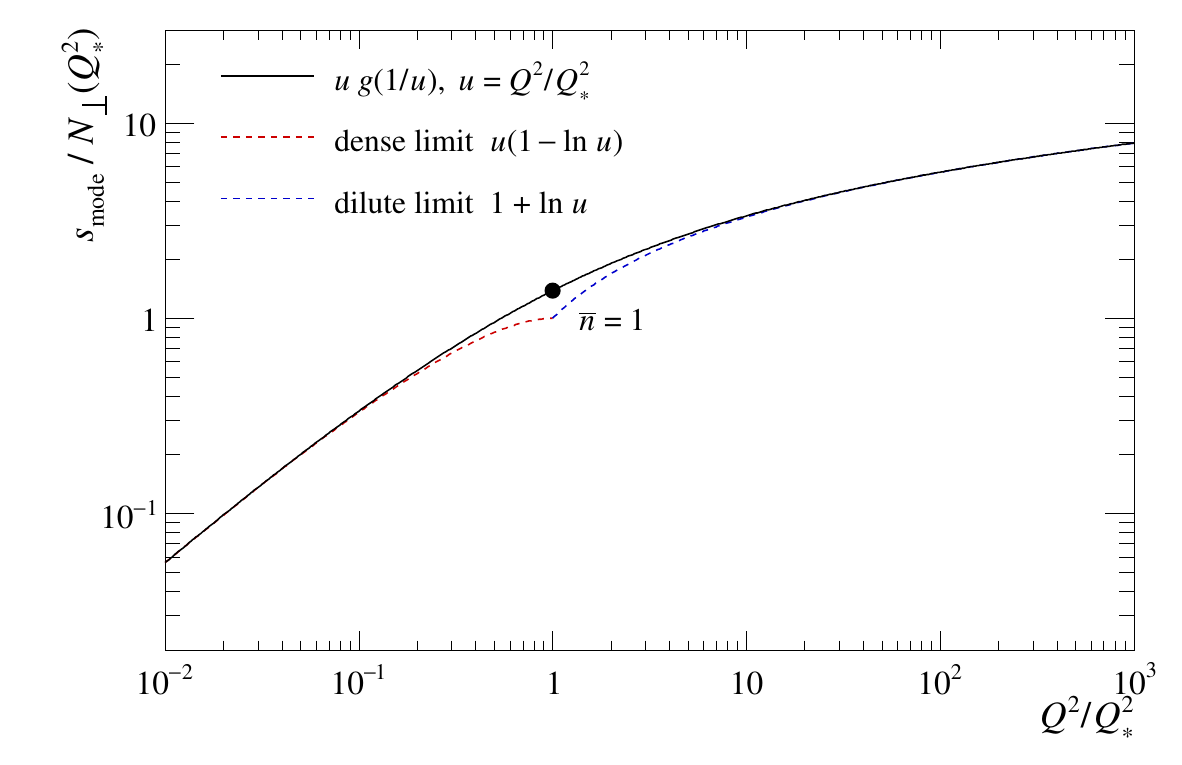}
\caption{Crossover of \eq{AC20} for one longitudinal cell per unit rapidity and gluon density independent of $Q^2$, fixed at its value at $Q_*$. Equal transverse occupations give the upper bound $s_{\rm mode}/N_\perp(Q_*^2)=u\,g(1/u)$ (solid), with $u=Q^2/Q_*^2$. The dense limit is $u(1-\ln u)$ (dashed red, $u\leq1$) and the dilute limit is $1+\ln u$ (dashed blue, $u\geq1$). The dot marks $\bar n=1$, where the curve equals $2\ln2$ and both limits give 1. At $\alpha_s=0.2$ the dense branch lies outside leading twist.}
\label{fig:crossover}
\end{figure}

The weight of Eq.~\eqref{B25} is $\alpha_s(1+n_{\rm shell})\simeq0.40$ at the occupancy crossover and $0.25$ at $Q_s$ for pure glue without quark pairs, $n_f=0$, fixed $\alpha_s=0.2$, $\gamma_G=0.3$, $N_c=3$ and $\varepsilon=10^{-2}$. The crossover is $Q_*^2\simeq0.15Q_s^2$.
 The gluon density independent of $Q^2$ used in Fig.~\ref{fig:crossover} instead gives $Q_*^2\simeq0.27Q_s^2$ at the same coupling. These estimates give the occupation and coupling dependence of \eq{B25}, not an error bound for a truncated vertex set.

Equal occupation of the transverse modes gives $s_{\rm mode}=N_\perp g(\bar n)$. Concavity of $g$ makes this an upper bound at fixed total gluon number. The dilute and dense limits are
\beq\label{AC20}
s_{\rm mode}\,\simeq\,xG\,\Bigl[1+\ln\frac{N_{\perp}}{xG}\Bigr]\quad (Q\gg Q_{*})\,,\qquad
s_{\rm mode}\,\simeq\,N_{\perp}\bigl[1+\ln\bar n\bigr]\quad (Q\ll Q_{*})\,.
\eeq
The dense expression holds for $\bar n\gg1$, or $Q^2\ll Q_*^2$ for a gluon density independent of $Q^2$, and the dilute expression holds for $\bar n\ll1$. 
The dilute entropy follows the gluon number, whereas the dense expression follows the transverse mode count up to its occupation logarithm.
 The proportionality to the transverse area follows from the extensivity of the mode count, not a boundary law of the kind reviewed by Eisert, Cramer and Plenio~\cite{EisertCramerPlenio}. Kovner and Lublinsky~\cite{KovnerLublinsky} found the same proportionality for the entanglement entropy of soft gluons in the Color Glass Condensate, which in their calculation is not extensive in rapidity. Both asymptotes are 28 percent low at $\bar n=1$. The entropy at that point is
\beq\label{AC20001}
s_{\rm mode}(Q_{*})\,=\,N_{\perp}(Q_{*})\,g_{B}(1)\,=\,\frac{N_{c}^{2}-1}{\pi}\,\ln2\;S_{\perp}Q_{*}^{2}\,,\qquad Q_{*}^{2}\,=\,c\,Q_{s}^{2}(x)/\alpha_{s}\,,
\eeq
where the relation to $Q_s$ assumes a gluon density independent of $Q^2$, and the small coefficient $c\simeq0.05$ enters together with the factor $1/\alpha_s$.

The crossover mode count is then $N_\perp(Q_*)=(N_c^2-1)S_\perp Q_s^2/(\pi N_c g_s^2)$, with $g_s^2=4\pi\alpha_s$. The infrared entropy term in Eq.~(3.21) of Kovner and Lublinsky~\cite{KovnerLublinsky} is $3(N_c^2-1)S_\perp Q_s^2/(8\pi g_s^2)$. Formally identifying the areas, couplings and squared saturation scales gives the ratio $8/(3N_c)\simeq0.9$ at $N_c=3$ between our mode count and that entropy term. The ratio for our crossover entropy has the additional factor $g_B(1)=2\ln2$. The two quantities and saturation-scale conventions differ, so this formal comparison does not fix their physical normalization.

Figure~\ref{fig:crossover} shows the change of slope for a gluon density independent of $Q^2$. The high-$Q$ entropy then grows logarithmically. The dilute entropy of \eq{AC20} grows with $xG$ when the gluon density increases with $Q^2$. 
The fixed-density result of Eq.~\eqref{AC19001001} puts the crossover below the saturation scale of \eq{AC19002} for $\alpha_s>c$, including $\alpha_s=0.2$, while $c/\alpha_s$ grows as the coupling decreases. The dense branch, $u\leq1$ in Fig.~\ref{fig:crossover}, therefore lies below the saturation scale for $\alpha_s>c$, where the collinear gluon density of \eq{AC190001} is outside the leading-twist domain.
 That part of the figure is a formal parametric estimate with an indicative overall scale. The one-mode limit $N_\perp=1$ removes the transverse counting and gives $g_B(xG)$. This equals \eq{AB26001} when $xG$ counts the emitted partons, $\bar n-1$, and agrees with it to relative order $1/(xG\ln xG)$ at large $xG$, as in Section~\ref{SecAOneMode}. Resolving $k_\perp$ interpolates between this limit and the extensive regime of \eq{AC20}. Quarks contribute at most $\ln2$ per mode, so their entropy has no logarithmic growth with the occupation. 
Their total has no fixed bound because the quark mode count also grows as $S_\perp Q^2$. The quark entropy stays small compared with the gluon entropy at small $x$ if the quark number is small compared with the gluon number and the gluon occupations are dilute, $Q\gg Q_*$. This suppression follows because the dilute mode entropy per parton, $1+\ln(1/n_\lambda)$, grows only logarithmically as the occupation falls. Small quark occupations alone do not make their entropy subleading.

\section{Conclusions}\label{SecConclusions}

We have shown that a one-parton density matrix constructed from Lipatov's parton wave functions separates the entropy of a sampled parton from the largest compatible state entropy, attained by a thermal state when it exists, when the resolved modes have finite mean number and finite entropy. The trace gives the average parton count, and dividing by that trace defines the sampled-parton state. 
The diagonal contains the leading-twist collinear densities and light-cone-gauge densities in transverse momentum.
 Helicity elements off the diagonal preserve phases between internal states of the same momentum. The mode entropy bounds the actual entropy and coincides with it only when the actual state is the thermal maximizer.
Entanglement further requires that the resolved modes form a subsystem of a pure joint state.
In the entanglement proposal~\cite{KharzeevLevin}, an entropy inferred from densities becomes a mode entropy only after the resolved modes and their finite traces and entropies are specified.
The geometric multiplicity law gives the one-mode value $g_B(\bar n-1)$, with $\bar n$ the mean multiplicity, if that mode contains every emitted parton.

We have also shown that averaging the parton state over unrecorded histories gives probability-preserving Lindblad evolution on a finite set of regulated modes in $t=\ln Q^2$, when different creation scales are treated as incoherent and the virtual loss obeys $\hat\Sigma=\tfrac12\hat{\mathcal R}^{\dagger}\hat{\mathcal R}$. This applies the known emission and no-emission description to collinear splitting on parton Fock space. 
The occupation evolution reduces to leading-order DGLAP, including virtual terms, at leading logarithmic accuracy as the cutoffs on momentum fractions are removed, for strongly ordered transverse momenta in light-cone gauge and dilute occupations, including those of newly resolved modes.
Each retained vertex set preserves the longitudinal momentum of the resolved modes in every state if its vertices conserve that momentum and its virtual partners are matched.
 Discarding coherence among distinct momentum and helicity modes and assigning thermal Gaussian populations gives a closed occupation equation and its mode-entropy production, with Bose and Pauli factors fixed by the operator algebra.

The identification of the light-cone-gauge densities in transverse momentum with time-reversal-even transverse-momentum distributions is not claimed beyond tree level or with final-state interactions (Sec.~\ref{SecAHelicity}). The approximations leave out momentum and color correlations beyond color-averaged thermal populations (Secs.~\ref{SecAMode} and~\ref{SecBOccupations}). The phase generator and soft-leg interference are also omitted (Secs.~\ref{SecBAmplitudes} and~\ref{SecBC}). The approximations also leave out an identification with the full coherent state (Sec.~\ref{SecBC}) and next-to-leading-order kernels, twist-four matching and recombination kernels (Sec.~\ref{SecBD}).
One open problem is to compute the entropy deficit $S_{\rm mode}-S_{\rm vN}$, the relative entropy to the thermal state with the same one-parton matrix. 
A calculation retaining the $1\to3$ vertex and connected pair cumulant $\hat C_2$ would compare the bound of Eq.~\eqref{B28} with the state entropy. It requires an independently specified truncation of the higher cumulants. The channels must be added as separate jumps, or the moments with unequal numbers of creation and annihilation operators of Appendix~\ref{AppD} must also be retained.

We propose comparing the mode number inferred from normalized hadron factorial moments in a rapidity window with the transverse mode count of Eq.~\eqref{AC19} for that window.
 The test would use deep inelastic scattering at HERA or proton-proton collisions at the LHC. The cluster ratio, the second factorial moment of one parton's hadron multiplicity divided by its squared mean, must be known with uncertainty small compared with the mean occupation per mode. Appendix~\ref{AppD} states the cluster and acceptance conditions.
Gelis, Lappi and McLerran~\cite{GelisLappiMcLerran} related the negative binomial parameter of produced gluon multiplicities to the number of color flux tubes in the glasma, the color field formed just after a nucleus-nucleus collision.
The saturation scale and transverse area set the glasma count, whereas our proposed comparison tests the inferred mode number, a negative binomial parameter, against the transverse mode count fixed by the resolution scale.
 Datta et al.~\cite{KLhadronization} tested the relation between the jet fragmentation function and the entropy of hadrons produced in jet fragmentation using ATLAS data at the LHC.

In conclusion, we have shown that Lipatov's wave functions give a one-parton density matrix separating sampled-parton entropy from mode entropy for modes of finite mean number and entropy.
 Mode entropy of such modes bounds state entropy if a thermal state with that matrix exists, equals it only for that state, and measures entanglement only if the resolved modes form a subsystem of a pure joint state. This matrix keeps leading-twist collinear densities, light-cone-gauge densities in transverse momentum and helicity phases at equal momentum. We have also shown that the state averaged over unrecorded emission histories, treated as incoherent at different scales, obeys Lindblad evolution on finitely many regulated modes with matched virtual terms. Its occupations reproduce leading-order DGLAP at leading logarithmic accuracy, including virtual terms, in light-cone gauge for dilute occupations, including those of newly resolved modes, and strongly ordered transverse momenta, as the cutoffs on momentum fractions are removed. Truncations keeping momentum-conserving vertices with matched virtual terms preserve, for any state, the longitudinal momentum of the resolved modes.
We have also shown that independent thermal populations of momentum and helicity modes close the occupation equation and give its mode-entropy production, with Bose and Pauli factors from the operator algebra.
We propose testing this thermal description by comparing the inferred and transverse mode counts.

\section*{Acknowledgments}
The authors used Claude Fable 5.1 and Claude Opus 5.5 (Anthropic) and GPT-6.1 Sol and GPT-6 Astra (OpenAI) to draft the text, to check derivations and numerical values by independent recomputation, and to write the programs that draw the figures. The authors specified each task, checked every derivation and number, and take full responsibility for the content.

\appendix
\section{Fock conventions and entropy details}\label{AppA}
\renewcommand{\theequation}{A.\arabic{equation}}
\setcounter{equation}{0}

\subsection{Packet normalization and the spectator trace}\label{AppAFock}

The states of identical partons include a symmetry factor for each species,
\beq\label{AA3001}
|n;\{\xi_i\}\rangle\,\equiv\,\prod_{s}\frac{1}{\sqrt{n_{s}!}}\,a^{\dagger}(\xi_1)\cdots a^{\dagger}(\xi_n)\,|0\rangle\,,
\qquad
\langle n;\{\xi'_i\}|n;\{\xi_i\}\rangle\,=\,\prod_{s}\frac{1}{n_{s}!}\sum_{\sigma_{s}}(\pm)^{\sigma_{s}}\prod_{i=1}^{n}
\delta(\xi_i-\xi'_{\sigma(i)})\,.
\eeq
Here $n_s$ counts the partons of species $s$, $\sigma_s$ permutes only identical partons, and $\sigma$ is the product of these permutations. The sign is positive for bosons and is the permutation signature for fermions. The corresponding symmetry or antisymmetry of $\psi_n$ cancels the permutation count in the norm.

A fixed-momentum hadron has normalization $\langle P'|P\rangle=16\pi^3P^+\delta^{(3)}(P'-P)$. The norm obtained from Eq.~\eqref{AA3} therefore contains $16\pi^3P^+\delta(0)\delta^{(2)}(0)$ multiplying $\sum_n\int[d\mu_n]|\psi_n|^2$. We use a normalizable packet $|h\rangle=\int dP^+d^2P_\perp\,\phi(P)|P\rangle$ and absolute parton momenta $k_i^+=x_i\bar P^+$ relative to its mean momentum. The packet smears the total longitudinal and transverse constraints. Fractions defined relative to the actual total momentum satisfy $\sum_i x_i=1$ identically, so the packet cannot smooth that constraint.

The total momentum variables are $X=\sum_i x_i$ and $K_\perp=\sum_i k_{\perp,i}$. The fraction $X$ is measured relative to the mean packet momentum $\bar P^+$. It equals one for a fixed-momentum hadron. The packet relation is $\Psi_n^{\rm pkt}(\{x_i,k_{\perp,i}\})=\Phi(X,K_\perp)\,X^{-(n-1)/2}\,\psi_n(\{\widetilde x_i,\widetilde k_{\perp,i}\})$. The intrinsic variables are $\widetilde x_i=x_i/X$ and $\widetilde k_{\perp,i}=k_{\perp,i}-\widetilde x_iK_\perp$.
 The intrinsic amplitude on the right is that of Eq.~\eqref{AA2}, with mass dimension $-(n-1)$. The profile $\Phi$ in the total packet momentum has mass dimension $-1$, so the canonical amplitude $\Psi_n^{\rm pkt}$ has mass dimension $-n$. 
The constrained measure $[d\mu_n]$ differs from the product measure by $16\pi^3\delta(1-\sum_i x_i)\delta^{(2)}(\sum_i k_{\perp,i})$. Smearing these total-momentum constraints introduces the packet profile. The symbol $\psi_n$ in Eq.~\eqref{AB7} and in the product-measure formulas below denotes $\Psi_n^{\rm pkt}$.

The change from $(x_i,k_{\perp,i})$ to $(X,K_\perp,\widetilde x_i,\widetilde k_{\perp,i})$ has Jacobian $X^{n-1}$. The factor $X^{-(n-1)/2}$ therefore makes each packet sector probability equal to the intrinsic sector probability, the corresponding term in Eq.~\eqref{AA4}, for any profile normalized by $\int dX\,d^2K_\perp\,|\Phi|^2/16\pi^3=1$. It also follows from smearing the Fock expansion of the fixed-momentum state $|P\rangle$, whose amplitudes are boost invariant. The distribution of total parton number is therefore unchanged at any packet width. Counts in fixed absolute cells, such as $x\geq x_{\rm min}$, change at second order in the width, as the density does.

The one-body density in absolute fractions is the intrinsic density smeared over the packet and approaches it as the packet narrows. The leading longitudinal correction is $\tfrac12\langle(X-1)^2\rangle\,\partial_x^2(x^2f)$ for a packet centered at $X=1$ and a density integrated over all transverse momenta. A transverse-momentum cut also receives a correction of the same order from the transverse packet spread.

Section~\ref{SecResolution} uses this packet in its translation-invariant limit, in which the packet becomes narrow in the hadron momentum and $\hat\rho_1$ is diagonal in $(x,k_\perp)$. The transverse area $S_\perp$ enters as a physical input. A normalizable packet of finite width can retain residual coherence between $x$ cells, measured by Eq.~\eqref{AC4}.

The single-parton measure is
\beq\label{AddAppA1}
\int d\xi \equiv \sum_{\rm f,c,s}\int dx\,d^2 k_\perp/16\pi^3
\eeq
and its identity kernel is
\beq\label{AB12}
\mathbb{1}(\xi;\xi')\,=\,16\pi^{3}\,\delta_{\rm fcs}\,\delta(x-x')\,\delta^{2}(k_{\perp}-k_{\perp}')\,,\qquad
\int d\xi'\;\mathbb{1}(\xi;\xi')\,f(\xi')\,=\,f(\xi)\,
\eeq
which fixes the normalization of the delta functions in the spectator formulas below. For one species the canonical packet has Fock expansion
\beq\label{eq:app-fock}
|h\rangle\,=\, \sum_n \frac{1}{\sqrt{n!}}\int \Big[\prod_{i=1}^{n} d\xi_i\Big]\;\psi_n(\xi_1,\dots,\xi_n)\,a^\dagger(\xi_1)\cdots a^\dagger(\xi_n)\,|0\rangle\,,
\eeq
with $\sum_n\int[\prod d\xi_i]|\psi_n|^2=1$. For several species the factorials and permutations separate as in Eq.~\eqref{AA3001}, and the factor attached to an observed parton counts its species.

The operator $a(\xi)$ contracts with each creation operator in an $n$-parton sector. The signs from the contractions cancel against the symmetry or antisymmetry of the amplitude, giving
\beq\label{eq:app-annihilated}
a(\xi)\,|h\rangle = \sum_n \frac{\sqrt{n}}{\sqrt{(n-1)!}}\int \Big[\prod_{i=2}^{n} d\xi_i\Big]\,\psi_n(\xi,\xi_2\dots\xi_n)\,
a^\dagger(\xi_2)\cdots a^\dagger(\xi_n)\,|0\rangle .
\eeq
The conjugate expression is
\beq\label{AddAppA2}
\langle h|\, a^\dagger(\xi') = \sum_m \frac{\sqrt{m}}{\sqrt{(m-1)!}}\int \Big[\prod_{j=2}^{m} d\eta_j\Big]\,\psi_m^{*}(\xi',\eta_2\dots\eta_m)\;
\langle 0|\, a(\eta_m)\cdots a(\eta_2)\,.
\eeq
Their product gives the density matrix before the spectator overlap is evaluated,
\beq\label{AB8001}
\hat\rho_1(\xi;\xi')\,=\,\sum_{n,m}\sqrt{nm}\int\prod_{i\geq2}d\xi_{i}\prod_{j\geq2}d\xi'_{j}\;
\bigl\langle{\rm spec}_{m-1};\{\xi'_{j}\}\bigl|{\rm spec}_{n-1};\{\xi_{i}\}\bigr\rangle\,
\psi_{n}(\xi,\{\xi_{i}\})\,\psi_{m}^{*}(\xi',\{\xi'_{j}\})\,,
\eeq
and the spectator states obey
\beq\label{AB8}
\bigl\langle {\rm spec}_{m-1};\{\xi_j'\}\bigl|{\rm spec}_{n-1};\{\xi_i\}\bigr\rangle
\,=\,\delta_{nm}\,\frac{1}{(n-1)!}\sum_{\sigma}(\pm)^{\sigma}\prod_{i\ge 2}\delta(\xi_i-\xi'_{\sigma(i)})\,,
\eeq
so only equal spectator numbers contribute. For $m=n$ the $(n-1)!$ contractions cancel the spectator normalization factors, leaving the factor $n$ in Eq.~\eqref{AB7}. If $P_n=\int[\prod d\xi_i]|\psi_n|^2$ is the sector probability, the trace is $\sum_n nP_n=\langle N\rangle$.

Hermiticity follows by conjugating Eq.~\eqref{AB6}. Positivity follows by choosing any square-integrable $u(\xi)$ and defining $\hat B=\int d\xi\,u^*(\xi)a(\xi)$,
\beq\label{AB8003}
\int d\xi\,d\xi'\;u^{*}(\xi)\,\hat\rho_1(\xi;\xi')\,u(\xi')
\,=\,\bigl\langle h\bigl|\hat B^{\dagger}\hat B\bigr|h\bigr\rangle
\,=\,\bigl\|\,\hat B\,|h\rangle\,\bigr\|^{2}\,\geq\,0\,.
\eeq

\subsection{Normalization and finite-cell entropy}\label{AppAEntropy}

The unnormalized one-parton matrix cannot be inserted into the von Neumann formula as a trace-one state. Substitution of $n_\lambda=\langle N\rangle p_\lambda$ gives
\beqar\label{AB13}
-\Tr\bigl[\hat\rho_1\ln\hat\rho_1\bigr]&=&-\sum_{\lambda}n_{\lambda}\ln n_{\lambda}\,=\,
-\sum_{\lambda}\langle N\rangle\,p_{\lambda}\Le \ln\langle N\rangle\,+\,\ln p_{\lambda}\Ra \nn \\
&=&\langle N\rangle\,\Bigl[\,S_{1}\,-\,\ln\langle N\rangle\,\Bigr]\,,\qquad
S_{1}\,\equiv\,-\sum_{\lambda}p_{\lambda}\ln p_{\lambda}\,.
\eeqar
All partons in one orbital have $S_1=0$, while $-\langle N\rangle\ln\langle N\rangle$ is negative for $\langle N\rangle>1$. 
An occupation $n_\lambda=100$ gives $-n_\lambda\ln n_\lambda=-460.5$, compared with $g_B(n_\lambda)=5.61$.
 The trace in Eq.~\eqref{AB13} includes the longitudinal variable and therefore has no remaining $x$ argument. 

The normalized slice $\varrho_x$ and finite transverse-cell probabilities $p_{j|x}$ of Section~\ref{SecAC} give
\beq\label{AB17}
S_{1}(x,Q^{2})\,=\,-\Tr\bigl[\varrho_{x}\ln\varrho_{x}\bigr]\,=\,
-\sum_j p_{j|x}\,\ln p_{j|x}\;+\;\sum_j p_{j|x}\,S\bigl(\varrho_{j,x}\bigr)\;-\;C_{\rm rel}\bigl(\varrho_{x}\bigr)\,.
\eeq
The matrix $\varrho_{j,x}$ is the normalized cell block defined in Section~\ref{SecAC}. The quantity $C_{\rm rel}$ is the transverse dephasing entropy of Eq.~\eqref{AB17002}.
The block entropy $S(\varrho_{j,x})$ need not vanish when the state is diagonal in the momentum cells.

The entropy of the full normalized matrix separates into longitudinal weights and slice entropies if $\hat\rho_1$ is block diagonal in the finite $x$ cells,
\beq\label{AB17001}
S_{1}\,=\,-\sum_i w_i\,\ln w_i\;+\;\sum_i w_i\,S_{1}(x_i,Q^{2})\,,\qquad
w_i\,=\,\int_{{\rm cell}\,i}dx\;f(x,Q^{2})/\langle N\rangle\,,
\eeq
where $\varrho_{x_i}$ is the normalized state in cell $i$ and $\sum_i w_i=1$. Insertion of Eq.~\eqref{AB17} and $W_{ij}=w_i p_{j|i}$ gives Eq.~\eqref{AB17003}, where longitudinal coherence and the continuum measure are discussed.

\subsection{The one-mode distribution and its dynamics}\label{AppAOneMode}

The shifted geometric distribution of Eq.~\eqref{AB23} solves the linear birth equation
\beq\label{AB27001}
\dot p_{n}=-n\lambda p_{n}+(n-1)\lambda p_{n-1}
\eeq
where the dot denotes differentiation in the cascade evolution variable and $\lambda=\omega$ of Eq.~\eqref{AB23} is its growth rate. Hentschinski, Kutak, P{\l}aczek and Rohrmoser~\cite{HKPRcascade} discuss this one-dimensional reduction of the dipole cascade, assuming a color average that leaves the state diagonal in dipole number. The entropy of that state is the number-distribution entropy. Its multiplicity entropy is Eq.~\eqref{AB26001}. In the leading-order small-$x$ identification, $\omega=4\ln2\,\bar\alpha_s$ is the intercept of the Balitsky, Fadin, Kuraev and Lipatov evolution~\cite{BFKL}, and is denoted $\Delta$ by Kharzeev and Levin~\cite{KharzeevLevin}.

The one-mode restriction of the parton splitting operator is $\hat{\mathcal R}=V a^{\dagger2}a$. Its matrix element is $\hat{\mathcal R}|n\rangle=Vn\sqrt{n+1}|n+1\rangle$, and the multiplicity probabilities obey
\beq\label{AB27003}
\dot p_{n}\,=\,|V|^{2}\,(n-1)^{2}n\,p_{n-1}\,-\,|V|^{2}\,n^{2}(n+1)\,p_{n}\,,
\eeq
with rates cubic in occupation, whereas those of Eq.~\eqref{AB27001} are linear.
Cheskis and Prygarin~\cite{CheskisPrygarin} write the pomeron cascade in zero transverse dimensions as $e^{\alpha Y(a^{\dagger2}a-a^\dagger a)}a^\dagger|0\rangle$, with splitting vertex $\alpha a^{\dagger2}a$ and propagation term $-\alpha a^\dagger a$, and obtain Eq.~\eqref{AB23} at $\omega=\alpha$ from the generating function $\langle0|e^{ua}e^{\alpha Y(a^{\dagger2}a-a^\dagger a)}a^\dagger|0\rangle$. 
They identify the state, at leading order in their power counting, with a squeezed and displaced one-parton state.
 The identity $\langle0|e^{ua}(a^\dagger)^n|0\rangle=u^n$ makes the probabilities coefficients of the unnormalized vectors $(a^\dagger)^n|0\rangle$, on which $a^{\dagger2}a(a^\dagger)^n|0\rangle=n(a^\dagger)^{n+1}|0\rangle$ gives the linear rates of Eq.~\eqref{AB27001}, while the same operator used as a jump on normalized Fock states in Eq.~\eqref{B12} gives the cubic rates of Eq.~\eqref{AB27003}.
 The shift $n=m+1$ in the linear birth equation gives
\beq\label{AB27004}
\dot p_{m}=\lambda[m\,p_{m-1}-(m+1)p_{m}]\,;
\eeq
the rate $\lambda(m+1)$ contains the stimulated contribution $\lambda m$ and spontaneous contribution $\lambda$. The dipole-counting factor $n-1$ is the Bose factor of the shifted mode.
In the density-operator evolution, Eq.~\eqref{AB27004} describes a one-leg gain process, while $a^{\dagger2}a$ has three legs in the same mode, so the equations differ in the number of legs and not in the statistics.

The factorization of Eq.~\eqref{B16} gives $n(1+n)^2$ when applied to one mode, compared with the number-state rate $n^2(n+1)$. The ratio of the latter to the factorized expression is $n/(1+n)$. The thermal value is $2n(n+1)(3n+1)$, as given after Eq.~\eqref{B16}. The ratio of the factorized expression to this thermal value tends to $1/6$ at large $n$ and is $0.168$ at $n=100$. This compares instantaneous rates in a thermal state. The exclusion of coincident daughter momenta used in deriving Eq.~\eqref{B16} has no effect at $N_\perp=1$, so it cannot justify that factorization here. The equality of the geometric entropy with the one-mode $S_{\rm mode}$, Eq.~\eqref{AB26001}, depends on the distribution and not on identifying these generators.

The factorized rate is 48 against the thermal value 240 for a geometric state of mean occupation 3 and $|V|^2=1$. The relative initial-slope error is 80 percent.
 This is a coincident-mode counting example. It gives neither a finite-time evolution nor an error estimate for distinct-mode collinear dynamics.

The many-mode expansion is
\beq\label{AB28}
S_{\rm mode}\,=\,\sum_{\lambda}\bigl(1+\ln n_{\lambda}\bigr)\,+\,O\!\Le \sum_{\lambda}\frac{1}{n_{\lambda}}\Ra
\,=\,M\,+\,\sum_{\lambda}\ln n_{\lambda}\,+\,O\!\Le \sum_{\lambda}\frac{1}{n_{\lambda}}\Ra\,,
\eeq
when every occupied gluon mode has $n_\lambda\gg1$, with $M$ the number of occupied modes. Equal occupations give Eq.~\eqref{AB28001}, and the ratio to $S_1$ is
\beq\label{AB28002}
\frac{S_{\rm mode}}{S_{1}}\,=\,\frac{M\bigl[1+\ln\langle N\rangle-\ln M\bigr]}{\ln M}
\,=\,\frac{\langle N\rangle}{\bar n}\,\frac{1+\ln\bar n}{\ln\bigl(\langle N\rangle/\bar n\bigr)}\,,
\eeq
where $\bar n=\langle N\rangle/M$. This expression requires $\bar n\gg1$ and gives a negative numerator if continued below $e^{-1}$. The dilute expression is instead
\beq\label{AB28004}
\langle N\rangle\bigl[1+\ln(1/\bar n)\bigr]/\ln\bigl(\langle N\rangle/\bar n\bigr)
\eeq
and approaches $\langle N\rangle$ logarithmically as $\bar n\to0$. 
The dilute expression of Eq.~\eqref{AB28004} gives 48 at $\bar n=10^{-1}$ and 80 at $\bar n=10^{-6}$ for $\langle N\rangle=100$.

\subsection{Asymptotic estimates for resolved cells}\label{AppAResolution}

The equal-bin example of Section~\ref{SecResolution} puts $M$ partons into one transverse mode and shares its occupation over $K$ longitudinal cells. The gap between the resolved and merged entropies has the dilute-cell limit
\beq\label{AC10001}
\Delta_{x}\,\simeq\,M\,\ln K\,+\,M\,\Bigl(1+\ln\frac{1}{M}\Bigr)\,-\,g(M)\,,
\eeq
for $M/K\ll1$. A decade of refinement then adds about $M\ln10\simeq2.3M$ nats.

The occupancy crossover follows from Eq.~\eqref{AC19001} by setting $\bar n(Q_*^2)=1$. The crossover is given by
\beq\label{AC19001001}
Q_{*}^{2}\,\simeq\, c\,Q_{s}^{2}(x)/\alpha_{s}\,,
\eeq
when the gluon density is independent of $Q^2$, so that the ratio in Eq.~\eqref{AC19001} is one. The power-law correction for a scale-dependent gluon density is given after Eq.~\eqref{AC19003}.

Subadditivity bounds the entropy increase when rapidity cells are specified as genuine subsystems of one joint state. In differential notation the bound is
\beq\label{AC21}
\frac{dS(Y)}{dY}\,\leq\,\sigma(Y)\,,
\eeq
where $S(Y)$ is the entropy of the union down to rapidity $Y$ and $\sigma(Y)$ is the isolated-slice entropy. The corresponding finite-cell statement bounds the entropy increment by the entropy of the added cell, with equality if and only if that cell is uncorrelated with the union. If the cells are subsystems of a pure cascade state, $S(Y)$ is the entanglement entropy of the cut at $Y$. This interpretation requires a subsystem partition, which is not supplied by the scalar multiplicity law alone.

The unit-cell profile of Eq.~\eqref{AC21001} gives, at large cell occupation,
\beq\label{AC21002}
\sigma(Y')\,\simeq\,\omega Y'\,+\,C\,;\qquad\, C\,=\,1\,+\,\ln(e^{\omega}-1)\,,
\eeq
with $C=0.29$ at $\omega\simeq0.4$. 
The cell occupation reaches one at $Y'\simeq1.8$, and the large-occupation approximation reaches five-percent accuracy only near $Y'\simeq5$, at occupation about 4.
 The ratio of the full cell entropy to this approximation is 3.3 at $Y'=0$. This is an asymptote for an individual cell. The sum over unit cells in Eq.~\eqref{AC22} has linear coefficient $C-\omega/2$, so its subleading term cannot be taken from the continuum integral of Eq.~\eqref{AC21002}.

\section{Leading-order DGLAP reduction}\label{AppB}
\renewcommand{\theequation}{B.\arabic{equation}}
\setcounter{equation}{0}

\subsection{Rate conventions}

We use the principal-value prescription for the light-cone gauge pole $1/(n\cdot k)$, as in Ref.~\cite{CFP} and the discussion by Bassetto, Nardelli and Soldati~\cite{BassettoNardelliSoldati}. 
The conversion from rate to Fock normalization after Eq.~\eqref{B6001} accounts for the transverse measure factor $1/16\pi^3$, the azimuthal factor $\pi$ and the longitudinal Jacobian $1/x_p$.
 The normalization of the ordered mode sums is
\beq\label{B9001}
V_{ab,c}\,=\,\frac{\mathcal V_{ab,c}}{S_{V}}\,,\qquad
V^{(k)}_{a_{1}\ldots a_{k+1},\,c}\,=\,\frac{\mathcal V^{(k)}_{a_{1}\ldots a_{k+1},\,c}}{S_{V}}\,,\qquad
S_{V}\,=\,(k+1)!\,,
\eeq
with $S_V=2$ for two-daughter channels. Each distinct final state appears $S_V$ times in an ordered sum and $S_V$ times again through the contractions. Every bilinear therefore has $S_V^2|V|^2=|\mathcal V|^2$, consistent with the normalized Fock-state matrix element given after Eq.~\eqref{B9}. The shell variable obeys
\beq\label{B9002}
k_{\perp}^{2}\,\delta\bigl(k_{\perp}^{2}-Q^{2}\bigr)\,=\,\delta\bigl(\lambda-t\bigr)\,,
\eeq
where $\lambda$ denotes $\ln k_\perp^2$ in this identity. The operator $\hat{\mathcal R}(t)$ acts at fixed radius after the scale delta function is integrated across the shell. Its squared matrix elements have the finite-rate normalization of Eq.~\eqref{B2}, giving for one parent
\beq\label{B9003}
\bigl\langle c\bigl|\hat{\mathcal R}^{\dagger}(t)\,\hat{\mathcal R}(t)\bigr|c\bigr\rangle\,=\,
\int_{0}^{1}\!dz\sum_{\rm spins,\,colors}\bigl|\mathcal V(z,k_{\perp})\bigr|^{2}k_{\perp}^{2}
\Big|_{k_{\perp}^{2}=e^{t}}\,=\,\int_{0}^{1}\!dz\;\frac{\alpha_{s}}{2\pi}\,\hat P^{(0)}(z)\,=\,
2\Sigma_{n}\,,
\eeq
where the integral counts distinct final states. The corresponding rates for quark and gluon parents are
\beqar\label{B5001}
\Sigma_{q}&=&\frac{1}{2}\int_{0}^{1}\!dz\;\frac{\alpha_{s}}{2\pi}\,\hat P^{(0)}_{qq}(z)\,, \nn \\
\Sigma_{g}&=&\frac{1}{2}\int_{0}^{1}\!dz\;\frac{\alpha_{s}}{2\pi}
\Bigl[\frac{1}{2}\hat P^{(0)}_{gg}(z)\,+\,n_{f}\,\hat P^{(0)}_{qg}(z)\Bigr]\,,
\eeqar
where the identical gluon pair has weight $1/2$, whereas the quark-pair term counts each flavor once.

The ratio of the symmetry factors of the two Fock sectors is
\beq\label{AppB6001001}
N_{n,n-1}\,=\,\Bigl[\,\prod_{s}\frac{n_{s}^{(n-1)}!}{n_{s}^{(n)}!}\,\Bigr]^{1/2},
\eeq
where $n_s^{(n)}$ counts the partons of species $s$ in sector $n$. The splitting sum in Eq.~\eqref{B6001} runs over unordered daughter pairs in the symmetrized Fock basis of Eq.~\eqref{AA3001}. Each term has this factor. It equals $1/\sqrt2$ for a gluon pair produced by a lone gluon and $1$ for distinct daughters of a lone parent. The gluon pair therefore has weight $1/2$ in Eq.~\eqref{B5001}.
 The channels $q\to qg$ and $q\to gq$ describe the same splitting with a different observed daughter, and the two daughter delta functions of Eq.~\eqref{B6001} supply their contributions.

We retain modes with $x\geq x_{\rm min}$ and impose the endpoint resolution $\varepsilon<z<1-\varepsilon$ on the splitting integrals. Jumps with either daughter outside the retained modes are omitted, and the virtual operator remains matched by Eq.~\eqref{B11}. A parent of fraction $x_p$ therefore has the range $\max(\varepsilon,x_{\rm min}/x_p)<z<1-\max(\varepsilon,x_{\rm min}/x_p)$. Its rate depends only on species for $x_p\geq x_{\rm min}/\varepsilon$, where the longitudinal cutoff does not act. These rates are given by Eq.~\eqref{B5001}, cut at the endpoint resolution. Below this threshold the range narrows and the rates depend on the parent fraction, vanishing for $x_p<2x_{\rm min}$.

This generator conserves the momentum of the retained modes identically. The occupations in the empty-daughter reduction obey the leading-order equations for observed fractions $x\geq x_{\rm min}/\varepsilon$, with every gain and loss integral cut to $\varepsilon<z<1-\varepsilon$. The endpoint logarithms cancel between gain and loss. Section~\ref{SecBDilute} states the limit. The finite cutoff also removes gain from parents above $x/\varepsilon$ for $x<\varepsilon$. This missing gain grows as the observed fraction falls below the endpoint cutoff. It is not suppressed by a power of that cutoff. Ordinary leading-order DGLAP densities, evolved without the cutoff and counted only above $x_{\rm min}$, lose momentum across the cutoff. The loss is 1.0 percent of the retained momentum per unit $\ln Q^2$ for a gluon input $xg\propto x^{-1/4}(1-x)^5$ at $\alpha_s=0.2$, $x_{\rm min}=10^{-3}$ and $n_f=0$.

Applying the shell operator to a common fixed set of modes gives the deterministic equation
\beq\label{B10}
Q^{2}\frac{\partial}{\partial Q^{2}}\,|\psi\rangle\,=\,\Bigl(-\hat\Sigma\,+\,\hat{\mathcal R}\Bigr)|\psi\rangle
\eeq
for a single deterministic amplitude. Its norm derivative for Hermitian $\hat\Sigma$ is
\beq\label{B10001}
Q^{2}\frac{\partial}{\partial Q^{2}}\langle\psi|\psi\rangle\,=\,
\bigl\langle\psi\bigr|\Bigl(-2\hat\Sigma\,+\,\hat{\mathcal R}\,+\,\hat{\mathcal R}^{\dagger}\Bigr)\bigl|\psi\bigr\rangle\,,
\eeq
and vanishing of this derivative for every state would require $2\hat\Sigma=\hat{\mathcal R}+\hat{\mathcal R}^\dagger$. No parton-number-conserving $\hat\Sigma$ satisfies this relation for a nonzero splitting operator, since the two sides connect different Fock sectors. The relation would also make the virtual operator first order in the vertex, whereas the Sudakov exponent is second order. This fixed-mode equation cannot conserve every state's norm. Equation~\eqref{B4} instead contains the boundary kernel that creates a fresh shell. Its amplitude solution obeys the integrated norm balance
\beq\label{B4001}
\sum_{n}\int[d\mu_{n}]\;\Sigma_{n}\,\bigl|\psi_{n}\bigr|^{2}\;=\;
\sum_{n}\int[d\mu_{n}]\;{\rm Re}\Bigl[\psi_{n}^{*}\,\mathcal R_{n,n-1}\,\psi_{n-1}\Bigr]\,,
\eeq
where components created at different scales are orthogonal. The prescription $\theta(0)=1/2$ fixes the boundary derivative below and makes the norm derivative vanish for the virtual loss of Eq.~\eqref{B11}. 

\subsection{Observed partons and spectator cancellation}

The amplitudes $\psi_n$ and probabilities $P_n$ use the packet and constrained-measure convention of Section~\ref{SecAWavefunctions} and Appendix~\ref{AppA}. The square of Eq.~\eqref{B4} obeys
\beq\label{AppB6003}
\frac{\partial|\psi_n|^{2}}{\partial t}\,=\,2\,{\rm Re}\Bigl[\psi_n^{*}\frac{\partial\psi_n}{\partial t}\Bigr]
\,=\,-\,2\,\Sigma_n\,\bigl|\psi_n\bigr|^{2}\;+\;
2\,{\rm Re}\Bigl[\psi_n^{*}\,\mathcal R_{n,n-1}\,\psi_{n-1}\Bigr]\,,
\eeq
and its contribution to the density is obtained from
\beq\label{AppB600301}
f_{i}(x,Q^{2})=\sum_{n}\int[d\mu_{n}]\sum_{l\in n}\delta(x-x_{l})\,\delta_{i,{\rm type}(l)}\,|\psi_{n}|^{2}\,.
\eeq
The virtual rate is the sum over the observed parton and all spectators. The observed contribution is
\beq\label{AppB6004}
-\,2\,\Sigma_{n}\Big|_{\rm observed}\;\longrightarrow\;-\,f_{i}(x,Q^{2})\int_{0}^{1}\!dz\;\frac{\alpha_{s}}{2\pi}\hat P^{(0)}(z)\,,
\eeq
where the parent kernels are defined in Section~\ref{SecBAmplitudes}.
Define $\sigma_{\tau}=2\Sigma_{\tau}$ for species $\tau$, with the cut range evaluated for each parent in the spectator sum.
A spectator splitting in sector $n-1$ gives
\beq\label{AppB600302}
A_{n-1}(x)\,=\,\int[d\mu_{n-1}]\;\delta(x-x_{l})\,\delta_{i,\tau(l)}
\sum_{p\neq l}\sigma_{\tau(p)}\,\bigl|\psi_{n-1}\bigr|^{2}\,,
\eeq
against the spectator loss $-A_n(x)$ in sector $n$. The sector sum telescopes if its boundary term vanishes, which holds at finite mean multiplicity away from $x=0$ for bounded spectator rates.
Every retained parton has $x\geq x_{\rm min}$ at finite cutoffs, so momentum conservation bounds the parton number of every sector by $1/x_{\rm min}$ and guarantees finite mean multiplicity.
 The constraint $\sum_i x_i=1$ in Eq.~\eqref{AA3} permits at most $1/x_0$ partons with $x_i\geq x_0>0$, so $\int_{x_0}^1 A_n(x)\,dx\leq\sigma_{\max}(n-1)P_n/x_0\to0$ when $\langle N\rangle$ is finite. We also require convergence permitting the sector sum and derivative to be interchanged. The stronger condition $\langle N(N-1)\rangle<\infty$ is needed here only for the spectator balance of the density integrated over all $x$, which gives the evolution of $\langle N\rangle$. It is not a condition on the norm balance~\eqref{B4001}.

The leading logarithmic iterate of the amplitude equation is
\beq\label{AppB600403}
\psi_{n}(t)\,=\,e^{-\int_{t_{0}}^{t}\Sigma_{n}\,dt''}\,\psi_{n}(t_{0})\,+\,
\int_{t_{0}}^{t}\!dt'\;e^{-\int_{t'}^{t}\Sigma_{n}\,dt''}\;
\mathcal R_{n,n-1}(t')\,\psi_{n-1}(t')\,,
\eeq
where each newly created component has the scale delta function of Eq.~\eqref{B6001}. The transverse and creation-scale integrations give the boundary derivative once in this sum of ordered emission amplitudes. At coincident scales we define that derivative by $\theta(0)=1/2$, which cancels the factor two in the real part of Eq.~\eqref{AppB6003}. This prescription introduces no squared distribution. The two-parton amplitude of Eq.~\eqref{eq:appb-psi2} gives the same unit coefficient without that prescription. It is nonzero only for $Q_0^2<k_\perp^2<Q^2$, and the derivative of its norm at the upper limit is the splitting rate. The initial component has transverse momenta $k_\perp\lesssim Q_0$, so its overlap with a newly created component is omitted at leading logarithmic accuracy. A kernel without locality in scale would give the growth $(\alpha_s t^2)^{n-1}$ in this iterate. The DGLAP growth is $(\alpha_s t)^{n-1}/(n-1)!$.

The two observed daughter terms yield
\beq\label{AppB6005}
2\,{\rm Re}\Bigl[\psi_n^{*}\,\mathcal R_{n,n-1}\,\psi_{n-1}\Bigr]\;\longrightarrow\;
\sum_{j}\int_{x}^{1}\frac{dz}{z}\;\frac{\alpha_{s}}{2\pi}\,\hat P^{(0)}_{ij}(z)\,f_{j}\!\Le\frac{x}{z},Q^{2}\Ra\,
\eeq
which is the real DGLAP convolution. The normalization follows from Eq.~\eqref{B2}, with the measure convention stated above. The cancellation with Eq.~\eqref{AppB6004} fixes the endpoint terms.

\subsection{Endpoint terms and sum rules}

The test function below treats the $z\to1$ endpoint. The lower gain cutoff is inactive only for $x\geq\varepsilon$. The endpoint cancellation uses
\beq\label{eq:dg-g}
g(z)\,\equiv\,\theta(z-x)\,\frac{1}{z}\,f\!\Le\frac{x}{z}\Ra\,,\qquad g(1)\,=\,f(x)\,.
\eeq
The quark loss integral is, up to terms of order $\varepsilon$,
\beq\label{eq:dg-q1}
\int_{0}^{1-\varepsilon}\!dz\;\frac{1+z^{2}}{1-z}\,\simeq\,2\ln\frac{1}{\varepsilon}\,-\,\frac{3}{2}\,,
\eeq
while the gain has the endpoint term $2g(1)\ln(1/\varepsilon)$ in addition to its plus-prescribed integral, up to terms of order $\varepsilon$ for $g$ smooth near $z=1$. The logarithms cancel between gain and loss. The limit $\varepsilon\to0$ gives
\beq\label{eq:dg-q3}
\hbox{gain}\,-\,\hbox{loss}\,=\,C_{F}\int_{0}^{1}\!dz\;
\Bigl[\frac{1+z^{2}}{(1-z)_{+}}\,+\,\frac{3}{2}\,\delta(1-z)\Bigr]\,g(z)\,,
\eeq
with quark endpoint coefficient $3C_F/2$.

The gluon kernel is symmetric under $z\to1-z$, so its loss integral satisfies
\beq\label{eq:dg-g1}
\frac{1}{2}\int_{\varepsilon}^{1-\varepsilon}\!dz\;\hat P^{(0)}_{gg}(z)\,=\,
\int_{1/2}^{1-\varepsilon}\!dz\;\hat P^{(0)}_{gg}(z)\,,
\eeq
where the integral of $\hat P^{(0)}_{gg}/2C_A$ over the half range is $\ln(1/\varepsilon)-11/12$, up to terms of order $\varepsilon$. The gluonic part therefore leaves the finite coefficient $11C_A/6$. The quark-pair loss subtracts $2n_fT_F/3$, giving
\beq\label{eq:dg-g4}
\frac{11C_{A}\,-\,4n_{f}T_{F}}{6}\,,
\eeq
so the regularized leading-order kernels are
\beqar\label{B2001}
P^{(0)}_{qq}(z)&=&C_{F}\Bigl[\frac{1+z^{2}}{(1-z)_{+}}\,+\,\frac{3}{2}\,\delta(1-z)\Bigr]\,, \nn \\
P^{(0)}_{gg}(z)&=&2C_{A}\Bigl[\frac{z}{(1-z)_{+}}+\frac{1-z}{z}+z(1-z)\Bigr]
\,+\,\frac{11C_{A}-4n_{f}T_{F}}{6}\,\delta(1-z)\,, \nn \\
P^{(0)}_{qg}(z)&=&T_{F}\Bigl[z^{2}+(1-z)^{2}\Bigr]\,, \nn \\
P^{(0)}_{gq}(z)&=&C_{F}\,\frac{1+(1-z)^{2}}{z}\,,
\eeqar
where the hats in Eq.~\eqref{B2} mean that the virtual delta terms are omitted and $(1-z)_+$ is replaced by $1-z$. The kernel $P^{(0)}_{qg}$ counts one flavor and one member of the pair. The total gluon decay rate contains $n_f$, while the singlet density and the momentum sum rule count both quarks and antiquarks.

The Altarelli and Parisi sum rules~\cite{AltarelliParisi} are
\beqar\label{B7}
\int_{0}^{1}dz\;P^{(0)}_{qq}(z)&=&0\,, \nn \\
\int_{0}^{1}dz\;z\,\Bigl[P^{(0)}_{qq}(z)+P^{(0)}_{gq}(z)\Bigr]&=&0\,, \nn \\
\int_{0}^{1}dz\;z\,\Bigl[2n_{f}P^{(0)}_{qg}(z)+P^{(0)}_{gg}(z)\Bigr]&=&0\,.
\eeqar
The first conserves quark number. The remaining two are the quark-parent and gluon-parent momentum sums. The kernels of Eq.~\eqref{B2001} satisfy these sum rules with the endpoint terms fixed above.

\subsection{Phase term, finite slabs and the entropy sign}

The phase contribution permitted by trace preservation has the form
\beq\label{B12001}
d\hat\rho/dt=-i[\hat H_{C},\hat\rho]+\hat{\mathcal R}\hat\rho\hat{\mathcal R}^{\dagger}
-\tfrac12\{\hat{\mathcal R}^{\dagger}\hat{\mathcal R},\hat\rho\}
\eeq
where the choice $\hat H_C=0$ and its effect on later occupations are stated in Section~\ref{SecBC}.

An initially empty shell mode with a prescribed constant source $w$ obeys the finite-slab equation
\beq\label{B17005}
\frac{dn_{\lambda}}{dt}\,=\,w\,\bigl(1+n_{\lambda}\bigr)\,,\qquad W\,=\,w\,\Delta t\qquad\Longrightarrow\qquad 1+n_{\lambda}(t^{+})\,=\,e^{W}\,,
\eeq
where $\Delta t$ is the slab width and $W=w\Delta t$ is the occupation produced without Bose enhancement. 
The finite-slab model holds $W$ fixed as $\Delta t\to0$, so its rate per mode $w=W/\Delta t$ diverges. Relating this prescription to the finite-rate normalization would require a shell-mode count proportional to $\Delta t$, which is not constructed here.

Equation~\eqref{B17005} retains one Bose factor and gives $n=W+W^2/2+\cdots$. Both daughters populate the shell. Equal daughter occupations with both Bose factors instead obey $dn/dt=w(1+n)^2$, giving $n=W/(1-W)=W+W^2+\cdots$ for an initially empty shell and $0\leq W<1$. 
Both enhancements are one power of the shell occupation relative to the linear result for $W=n_{\rm shell}\ll1$. Their coefficients lie beyond leading logarithmic accuracy. The shell is empty before the last splitting at this accuracy, and the leading logarithmic kernel is unchanged in this dilute-shell regime.

The source is identified with the new-shell occupation $n_{\rm shell}$ evaluated in Section~\ref{SecResolution} and used in Eq.~\eqref{B25}. The pure-gluon example has no quark pairs, $n_f=0$, fixed coupling $\alpha_s=0.2$, $\gamma_G=0.3$ and $N_c=3$. At its occupancy crossover, $W$ reaches 1 for $\varepsilon\lesssim10^{-2}$. The dilute expansion then fails, and the two-factor form $W/(1-W)$ has no finite value.

The entropy sign in Eq.~\eqref{B17002} concerns dimensionless cell occupations. A longitudinal cell of fixed rapidity width $\Delta Y$ has occupation proportional to $\Delta Y\,xG(x)$, whereas a cell of fixed width $\Delta x$ has occupation proportional to $\Delta x\,G(x)$, for the same transverse mode count. The cell measure and its vertex normalization must be kept consistent. Consider $n\propto x^{-\lambda}$ for one specified measure and a common transverse prefactor for the parent and daughters, with $x_a=zx_c$ and $x_b=(1-z)x_c$. Here $\lambda$ is the longitudinal power-law exponent. 
The large-occupation condition is
\beq\label{B17004}
z^{\lambda}\,+\,(1-z)^{\lambda}\,>\,1\,,
\eeq
which holds for $0<z<1$ and $\lambda<1$. 
The per-channel criterion for nonnegative entropy flow stated in Section~\ref{SecBEntropy} holds for this profile at every finite occupation when $\lambda\leq1$.
 At $\lambda=1$, the entropy weight is $\epsilon_a+\epsilon_b-\epsilon_c=\ln[1+z(1-z)/(n_c(n_c+1))]>0$. For $\lambda>1$, the sign becomes negative once $n_c$ exceeds $[z(1-z)]^\lambda/[1-z^\lambda-(1-z)^\lambda]$.

The exponent $\lambda\simeq0.29$ of Golec-Biernat and W\"usthoff~\cite{GBW} gives the dilute-model example for cells of fixed rapidity width, as in Section~\ref{SecResolution}. The same longitudinal density gives exponent $1.29$ for cells of fixed width in $x$. Its entropy weight at $z=1/2$ becomes negative above $n_c\simeq0.918$. Thus the positive-sign example for rapidity cells does not establish the sign for a different cell measure.

\section{Helicity amplitudes and ordered emissions}\label{AppC}
\renewcommand{\theequation}{C.\arabic{equation}}
\setcounter{equation}{0}

The amplitude ansatz \eq{B4} can be solved successively for a quark and its first emissions. The initial state is one quark, not a color singlet hadron. 
The amplitudes below use the rate convention of Eq.~\eqref{B2}, whose squared vertex determines the real splitting probability. The initial quark has $x_p=1$, so its two-body Fock amplitude for the constrained measure~\eqref{AA3} is $4\pi$ times the rate amplitude. Each later emission has the factor $4\pi/\sqrt{x_p}$ obtained after the delta functions in Eq.~\eqref{B6001} are evaluated.

The one-parton sector has only virtual evolution,
\beq\label{eq:appb-psi1eq}
\frac{\partial\psi_{1}}{\partial t}\,=\,-\,\Sigma_{1}\,\psi_{1}\,,\qquad t\,=\,\ln Q^{2}\,.
\eeq
Its intrinsic amplitude is $e^{-I}$, where $I\bigl(Q_0^2,Q^2\bigr)=\int_{Q_0^2}^{Q^2}(dq^2/q^2)\,\Sigma_1\bigl(q^2\bigr)$. The longitudinal and transverse delta functions enforcing one-parton kinematics belong to the measure \eq{AA3}. The square of this amplitude is the Sudakov no-emission probability,
\beq\label{eq:appb-sud}
\Delta\bigl(Q_{0}^{2},Q^{2}\bigr)\,=\,e^{-2I}\,,\qquad
\bigl|\psi_{1}\bigr|^{2}\,\propto\,\Delta\,.
\eeq

An emission supplies the source for the two-parton sector. Momentum conservation imposes
\beq\label{eq:appb-kin}
x_{b}+x_{c}\,=\,1\,,\qquad k_{\perp b}+k_{\perp c}\,=\,0\,,
\eeq
so its independent variables are $z\equiv x_b$ and $k_\perp\equiv k_{\perp b}$. The scale delta function in \eq{B6001} fixes the emission at $t=\ln k_\perp^2$. 
The integrated source gives, in the rate convention,
\beq\label{eq:appb-psi2}
\psi^{\lambda'\lambda_{g}}_{2}\bigl(z,k_{\perp};Q^{2}\bigr)\,=\,
\sqrt{\Delta_{1}\bigl(Q_{0}^{2},k_{\perp}^{2}\bigr)}\;
\mathcal V^{\lambda'\lambda_{g}}_{\lambda}\bigl(z,k_{\perp}\bigr)\;
\sqrt{\Delta_{2}\bigl(k_{\perp}^{2},Q^{2}\bigr)}\,.
\eeq
The no-emission probability $\Delta_1$ describes the quark parent before the splitting. The probability $\Delta_2$ is the product of the quark and gluon no-emission probabilities afterward. For $q\to qg$, the quark keeps the fraction $z$ and the vertex is
\beq\label{eq:appb-vertex}
\mathcal V^{\lambda'\lambda_{g}}_{\lambda}\bigl(z,k_{\perp}\bigr)\,=\,g_{s}\,T^{a}\,
\frac{\epsilon^{*}_{\lambda_{g}}\!\cdot k_{\perp}}{k_{\perp}^{2}}\;
C^{\lambda'\lambda_{g}}_{\lambda}(z)\,,
\eeq
where we choose the transverse polarization phases as
\beq\label{eq:appb-pol}
\epsilon^{*}_{\pm}\!\cdot k_{\perp}\,=\,\mp\,\frac{|k_{\perp}|}{\sqrt{2}}\;e^{\mp i\varphi}\,,
\eeq
with $\varphi$ the azimuthal angle. The quark helicity is conserved for massless quarks, $\lambda'=\lambda$, and the relative gluon helicities give
\beq\label{eq:appb-C}
C^{\lambda,\,\lambda_{g}=\lambda}_{\lambda}(z)\,=\,\frac{1}{2\pi}\,\frac{1}{\sqrt{1-z}}\,,\qquad
C^{\lambda,\,\lambda_{g}=-\lambda}_{\lambda}(z)\,=\,\frac{1}{2\pi}\,\frac{z}{\sqrt{1-z}}\,,
\eeq
whose magnitudes obey parity. The two helicity contributions to the real kernel are $C_F/z_g$ and $C_F(1-z_g)^2/z_g$, with $z_g=1-z$ the gluon fraction. Their sum is $P^{(0)}_{gq}$ and their difference is the polarized kernel $\Delta P^{(0)}_{gq}(z_g)=C_F(2-z_g)$.

The spin and color sum, with the parent average understood, gives
\beq\label{eq:appb-norm}
\sum_{\lambda'\lambda_{g}}\bigl|\mathcal V\bigr|^{2}\,=\,
g_{s}^{2}C_{F}\,\frac{k_{\perp}^{2}/2}{k_{\perp}^{4}}\,\frac{1}{4\pi^{2}}\,\frac{1+z^{2}}{1-z}
\,=\,\frac{\alpha_{s}}{2\pi}\,\frac{\hat P^{(0)}_{qq}(z)}{k_{\perp}^{2}}\,,
\eeq
which agrees with \eq{B2}.
The overall phase is a spinor normalization convention. The relative sign of the two helicity coefficients is fixed and positive once Eq.~\eqref{eq:appb-pol} is chosen. This sign and the azimuthal phases give the off-diagonal gluon-helicity elements in \eq{eq:appb-rhohel}. A square root of the inclusive kernel does not retain them. 

The trace-normalized helicity block of Sec.~\ref{SecA} has eigenvalues and coherence
\beq\label{eq:appb-lam}
\lambda_{\pm}\,=\,\frac{(1\pm z)^{2}}{2(1+z^{2})}\,,\qquad
C_{\rm hel}(z)\,=\,\ln2\,+\,\sum_{\pm}\lambda_{\pm}\ln\lambda_{\pm}\,=\,
\frac{(1+z)^{2}\ln(1+z)+(1-z)^{2}\ln(1-z)}{1+z^{2}}\,-\,\ln(1+z^{2})\,,
\eeq
where the coherence $C_{\rm hel}$ is defined relative to removal of the off-diagonal helicity elements. It runs from $0$ at $z\to0$ to $\ln2$ at $z\to1$, with values $0.020$, $0.368$, and $0.674$ at $z=0.1$, $0.5$, and $0.9$. The internal entropy of this block is $\ln2-C_{\rm hel}(z)$, which vanishes as $z\to1$. The assumptions of the helicity example and its relation to the gluon distribution are stated in Section~\ref{SecAHelicity}.

We use the plus component of Lepage and Brodsky~\cite{LepageBrodsky} and $k^-=(k_\perp^2+m^2)/k^+$. The massless two-body invariant energy denominator is $-k_\perp^2/[z(1-z)]$, independent of the light-cone convention. This denominator is real and negative below threshold, so the tree-level wave function has no phase beyond the azimuthal phases of the vertices, consistent with Section~\ref{SecB}. Its factor $z(1-z)$ combines with the numerator and the vertex normalization to give \eq{eq:appb-C}. 
Equation~(3.79) of Ref.~\cite{BrodskyPauliPinsky}, for the corresponding abelian channel, agrees with the helicity-conserving component of the undressed vertex~\eqref{eq:appb-vertex} for massless fermions and opposite boson and fermion helicities, including the azimuthal phase, after conversion to the Fock normalization and from the boson to the quark momentum fraction.
 Squaring the rate amplitude gives
\beq\label{eq:appb-psi2sq}
\bigl|\psi_{2}\bigr|^{2}\,=\,\frac{\alpha_{s}}{2\pi}\,\frac{\hat P^{(0)}(z)}{k_{\perp}^{2}}\;
\Delta_{1}\bigl(Q_{0}^{2},k_{\perp}^{2}\bigr)\;\Delta_{2}\bigl(k_{\perp}^{2},Q^{2}\bigr)\,,
\eeq
which is the radial splitting probability density between two no-emission probabilities.
The amplitude ansatz does not reproduce the two-loop light-cone calculation of the nonsinglet splitting function in Ref.~\cite{LappiNLOLCPT}.

The boundary amplitude equation \eq{B4} has the ordered solution
\beq\label{eq:appb-U}
\bigl|\psi(t)\bigr\rangle\,=\,U\bigl(t,t_{0}\bigr)\,\bigl|\psi(t_{0})\bigr\rangle\,,\qquad
U\bigl(t,t_{0}\bigr)\,=\,{\rm P}\exp\Bigl[-\int_{t_{0}}^{t}dt'\,
\bigl(\hat\Sigma(t')-\hat{\mathcal K}(t')\bigr)\Bigr]\,,
\eeq
where $\hat{\mathcal K}(t)$ denotes the boundary kernels $\mathcal R_{n,n-1}$ of Eq.~\eqref{B6001} acting on all sectors, including their scale delta function. Expanding in the boundary kernel while retaining virtual evolution gives
\beq\label{eq:appb-dyson}
U\,=\,U_{0}\bigl(t,t_{0}\bigr)\,+\,\int_{t_{0}}^{t}\!dt_{1}\;
U_{0}\bigl(t,t_{1}\bigr)\,\hat{\mathcal K}\bigl(t_{1}\bigr)\,U_{0}\bigl(t_{1},t_{0}\bigr)\,+\,\ldots\,,
\qquad
U_{0}\bigl(t,t'\bigr)\,=\,\mathrm{P}\,e^{-\int_{t'}^{t}\hat\Sigma}\,.
\eeq
The virtual operator is a sum of commuting rates fixed by the parton species for empty daughter modes and parents with $x_p\geq x_{\rm min}/\varepsilon$. The rates are those of Eq.~\eqref{B5001}, cut at the endpoint resolution in Appendix~\ref{AppB}, so the ordering in $U_0$ can be omitted.
 Diagonality in parton number alone does not imply commutation at different scales. For one $1\to2$ vertex, $n$ emission insertions give the $(n+1)$-parton amplitude. 
The first insertion gives \eq{eq:appb-psi2} in the rate convention after the scale delta function collapses its integration.

The three-parton amplitude can contain two ordered emissions or one $1\to3$ vertex,
\beq\label{eq:appb-psi3}
\psi_{3}\;\sim\;\underbrace{\mathcal V^{(1)}\mathcal V^{(1)}}_{\hbox{two ordered $1\to2$ splittings}}
\;+\;\underbrace{\mathcal V^{(2)}}_{\hbox{one $1\to3$ vertex}}\,,
\eeq
where the superscripts on $\mathcal V^{(1)}$ and $\mathcal V^{(2)}$ count the increase in parton number, not loops.
Their probability contributions have the following logarithmic weights at fixed coupling, with the two scale integrations ordered as $k_{\perp,1}^2<k_{\perp,2}^2$
\beq\label{eq:appb-logs}
\hbox{two ordered vertices}\;\sim\;\frac{\alpha_{s}^{2}}{2}\,\ln^{2}\frac{Q^{2}}{Q_{0}^{2}}\,,\qquad
\hbox{one $1\to3$ vertex}\;\sim\;\alpha_{s}^{2}\,\ln\frac{Q^{2}}{Q_{0}^{2}}\,,
\eeq
so the ordered emissions dominate at leading logarithmic accuracy. The $1\to3$ vertex, after subtraction of the iterated $1\to2$ contribution, enters at the order counted in \eq{B27}.

\section{Moment hierarchy and coherent splitting channels}\label{AppD}
\renewcommand{\theequation}{D.\arabic{equation}}
\setcounter{equation}{0}

\subsection{Reduced operators and multiplicity moments}

The one-body operator belongs to a family that measures correlations among several observed partons. For a pure state, its rank-$m$ member is
\beq\label{AppE1}
\hat\rho_m\bigl(\xi_1\ldots\xi_m;\,\xi'_1\ldots\xi'_m\bigr)\,=\,
\bigl\langle h\bigr|\,a^{\dagger}(\xi'_m)\cdots a^{\dagger}(\xi'_1)\,
a(\xi_1)\cdots a(\xi_m)\,\bigl|h\bigr\rangle\,,
\eeq
where the composite label $\xi$ is defined in Eq.~\eqref{AB4}. For the history-averaged state, the expectation value is taken with $\hat\rho$. The Fock expansion gives
\beq\label{AppE2}
\hat\rho_m\bigl(\xi_1\ldots\xi_m;\,\xi'_1\ldots\xi'_m\bigr)\,=\,
\sum_{n\geq m}\frac{n!}{(n-m)!}\int\prod_{i=m+1}^{n}d\xi_i\;
\psi_n\bigl(\xi_1\ldots\xi_m,\{\xi_i\}\bigr)\,
\psi^{*}_n\bigl(\xi'_1\ldots\xi'_m,\{\xi_i\}\bigr)\,,
\eeq
where the factor $n!/(n-m)!$ counts ordered choices of the observed partons. For several species it becomes the product of the falling factorials of the observed species. The spectators are integrated with their full momentum, spin, color and flavor measure.
 Each operator has equal numbers of creation and annihilation operators, so it preserves parton number and does not measure coherence between different Fock sectors. Its trace is a factorial moment,
\beq\label{AppE3}
\Tr\hat\rho_m\,=\,\int d\xi_1\cdots d\xi_m\;\hat\rho_m\bigl(\xi_1\ldots\xi_m;\,\xi_1\ldots\xi_m\bigr)
\,=\,\bigl\langle N(N-1)\cdots(N-m+1)\bigr\rangle\,,
\eeq
which is finite when the corresponding multiplicity moment exists. The diagonal elements of $\hat\rho_m$ in a mode basis are the normal-ordered occupation correlators $\langle{:}N_{\lambda_1}\cdots N_{\lambda_m}{:}\rangle$. For $m=1$, they are the occupations $n_\lambda$. The connected mode cumulants used below are formed from these correlators.

The factorial moments provide a test of the independent geometric mode model. The total multiplicity of $M$ equally occupied bosonic modes with mean $\bar n$ per mode has mean $\mu=M\bar n$ and variance $V=M\bar n(1+\bar n)$. Thus $M=\mu^2/(V-\mu)$ and $\bar n=(V-\mu)/\mu$ for $\mu>0$ and $V>\mu$, with $M$ a positive integer if interpreted as a literal number of modes. The total multiplicity is negative binomial with parameter $k=M$. Its inversion from the first two moments is the relation used by Kutak and Prasza{\l}owicz~\cite{KutakPraszalowicz}.
A negative binomial also arises for gluons produced at central rapidity in the glasma of a nucleus-nucleus collision, with both nuclei treated as Color Glass Condensates~\cite{GelisLappiMcLerran}.
That result holds at leading logarithmic accuracy in $1/x$ and lowest order in the color sources, with infrared divergences cut off at the saturation scale.
There $k$, proportional to $(N_c^2-1)Q_s^2S_\perp$ with an undetermined coefficient of order one, is an effective count of independent emitters with geometric counts associated with glasma color flux tubes.
Here $k=M$ counts transverse parton modes resolved at $Q$.

At fixed $M$, the normalized factorial moments $\Gamma(M+q)/[\Gamma(M)M^q]$ are independent of $\mu$, and as $\mu\to\infty$ the KNO scaling function is $M^M\zeta^{M-1}e^{-M\zeta}/\Gamma(M)$ for $\zeta=n/\mu$, reducing to $e^{-\zeta}$ at $M=1$.
 
Ouchen and Prygarin~\cite{OuchenPrygarinKNO} expanded the shifted geometric law at large mean to obtain inverse-mean corrections to its KNO scaling function $e^{-\zeta}$. The first correction vanishes at $\zeta=2$. There the KNO-scaled distributions measured by ATLAS for charged hadrons with transverse momentum above 500 MeV lie within five percent of $e^{-2}$~\cite{OuchenPrygarinKNO}.

Ouchen and Prygarin~\cite{OuchenPrygarinRecip} derived a relation between the multiplicity distribution at its mean and its derivative with respect to multiplicity there. The relation is $P'(\mu)=-P(\mu)/\mu$, where $P$ is the multiplicity distribution and the prime denotes differentiation with respect to $n$. The slope follows from the approximate reciprocal symmetry $f_s(\zeta)=f_s(1/\zeta)$ of the relative deviation (Sec.~\ref{SecAOneMode}). This symmetry was reported for $1/3<\zeta<3$ in ATLAS and CMS charged-hadron data at 7 to 13 TeV~\cite{OuchenPrygarinRecip} and not established at 13 TeV~\cite{OuchenPrygarinSandeu}. The slope relation holds for every negative-binomial shape $k$ in the KNO limit~\cite{OuchenPrygarinSandeu}. It therefore cannot determine $M=k$ in our mode model. At finite mean our model satisfies the relation up to relative corrections of order $M/\mu$.

 The model then predicts the higher factorial cumulants, $\kappa_r^{(F)}=(r-1)!M\bar n^r$. Unequal occupations leave the higher cumulants undetermined by the first two moments. The entropy of the vector of mode occupations is $Mg_B(\bar n)$, whereas the entropy of the total multiplicity also loses the information about how a given count is distributed among the modes.

A stipulated example with $M=2$ and $\bar n=3$ has $\mu=6$, $V=24$ and $\kappa_3^{(F)}=108$. These are model values, not measured multiplicities.

A test with hadron multiplicities requires a specified cluster law and acceptance. Normalized factorial moments at fixed order approach their partonic values, with inverse-$\mu$ corrections, for independent clusters drawn from an energy-independent law with positive mean and finite moments through that order, provided the normalized parton moments remain bounded.

Independent clusters confined to the window of their parent parton add $r_{\rm cl}/\mu$ to the second normalized factorial moment~\cite{AdelajaPrygarinYusuf}.
 Here $r_{\rm cl}$ is the second factorial moment of the cluster multiplicity divided by the square of its mean. The fitted value is $M_{\rm fit}=M/(1+r_{\rm cl}M/\mu)$, so its bias depends on the mean occupation per mode $\mu/M$. The cluster ratio must therefore be known with an uncertainty small compared with this occupation to infer the partonic mode count. The fit returns 35 percent of $M$ for two fixed hadrons with $r_{\rm cl}=1/2$ and 12 percent for geometric clusters with $r_{\rm cl}=2$. Both values use $\mu/M=0.27$, the occupation at $Q_s$ in Section~\ref{SecResolution}.

Uniform independent thinning keeps each particle independently with the same probability and preserves the geometric family. Mode-dependent or correlated acceptance requires its own treatment. Unnormalized factorial moments retain the cluster factors, so hadron mean and variance cannot in general be substituted into the partonic formulas for $M$ and $\bar n$. A negative-binomial fit alone does not establish a thermal parton state.

The extension to normalized factorial moments adds a correlation assumption to the local parton-hadron duality of Azimov, Dokshitzer, Khoze and Troyan~\cite{LPHD}, formulated for inclusive spectra. Dremin and Gary~\cite{DreminGary} review evidence for that duality in full phase space. Its extension to correlations in restricted phase-space bins is not established, as discussed by De Wolf, Dremin and Kittel~\cite{DDKreview}.

The connected part of the two-body operator is defined by
\beq\label{AppE4}
\hat\rho_2\bigl(\xi_1\xi_2;\xi'_1\xi'_2\bigr)\,=\,
\hat\rho_1\bigl(\xi_1;\xi'_1\bigr)\hat\rho_1\bigl(\xi_2;\xi'_2\bigr)
\,\pm\,\hat\rho_1\bigl(\xi_1;\xi'_2\bigr)\hat\rho_1\bigl(\xi_2;\xi'_1\bigr)
\,+\,\hat C_2\bigl(\xi_1\xi_2;\xi'_1\xi'_2\bigr)\,,
\eeq
where the upper sign applies to gluons and the lower sign to quarks. The exchange term accounts for statistics. Higher connected operators are defined recursively by
\beq\label{AppE5}
\hat\rho_m\,=\,\sum_{\rm pairings}\bigl(\pm\bigr)^{P}\prod\hat\rho_1
\;+\;\hbox{terms containing }\hat C_2
\;+\;\cdots\;+\;\hat C_m\,,
\eeq
where the sign $(\pm)^P$ is the permutation signature for fermions and unity for bosons. A thermal Gaussian state has every $\hat C_m=0$ for $m\geq2$, as imposed in Eq.~\eqref{B15001}.  

\subsection{The evolution hierarchy}

Let $\hat O_m$ be the operator string in Eq.~\eqref{AppE1}. The adjoint of the coherent density evolution~\eqref{B12}, with $\hat{\mathcal R}=\sum_k\hat{\mathcal R}_k$, gives
\beq\label{AppE6}
\frac{d\hat\rho_m}{dt}\,=\,\frac{1}{2}\sum_{k,k'\geq1}
\Bigl[\,\Bigl\langle\,\hat{\mathcal R}^{\dagger}_{k'}\,\bigl[\hat O_m,\hat{\mathcal R}_{k}\bigr]\,\Bigr\rangle
\,+\,\Bigl\langle\,\bigl[\hat{\mathcal R}^{\dagger}_{k'},\hat O_m\bigr]\,\hat{\mathcal R}_{k}\,\Bigr\rangle\,\Bigr]\,.
\eeq
On the diagonal, $\hat O_m$ is Hermitian and the two terms become conjugates after exchange of $k,k'$. Both terms are needed off the diagonal. For $k\neq k'$, normal ordering gives moments with unequal numbers of creation and annihilation operators. The number-conserving family $\hat\rho_m$ must then be enlarged by these anomalous moments.

For $k=k'$, normal ordering gives the schematic range
\beq\label{AppE7}
\frac{d\hat\rho_m}{dt}\Big|_{k=k'}\,=\,\sum_{k\geq1}\;\sum_{r=1}^{m+k+1}\,
\mathcal W^{(k)}_{m\leftarrow r}\bigl[\hat\rho_{r}\bigr]\,,
\eeq
where $\mathcal W^{(k)}_{m\leftarrow r}$ is the linear map from rank $r$ to rank $m$ built from $|V^{(k)}|^2$. Some ranks have zero coefficients. Ranks below $m$ occur because one splitting can create a correlated pair from a single parent. The hierarchy of Bogoliubov, Born, Green, Kirkwood and Yvon~\cite{BBGKY} therefore contains sources from both lower and higher moments, all fixed by the supplied splitting vertices.

The double-parton evolution of Kirschner and of Shelest, Snigirev and Zinovjev~\cite{DoubleDGLAP} has a source from one-parton densities. Diehl, Ostermeier and Sch\"afer~\cite{DoubleDGLAP} give homogeneous evolution at nonzero transverse separation. The source appears after integration over that separation and ultraviolet subtraction.

The three-body virtual operator reduces to commuting rates when strong ordering supplies empty daughter modes, as described after Eq.~\eqref{B11}.
 Truncating the vertex set at $k$ does not limit the connected cumulants to $\hat C_2,\ldots,\hat C_{k+1}$. A single $1\to2$ vertex already generates a connected three-mode cumulant from a thermal parent. Setting $\hat C_m=0$ above a chosen rank is a separate approximation, with no identification between cumulant rank and QCD twist supplied by the Fock grading.

\subsection{Coherent operators and parity}

Two different multiplicity changes lead to orthogonal final sectors for an initial state of definite parton number,
\beq\label{B19002}
\bigl\langle\psi\bigl|\hat{\mathcal R}^{\dagger}_{k}\hat{\mathcal R}_{k'}\bigr|\psi\bigr\rangle\,=\,0
\qquad\hbox{for every $|\psi\rangle$ of definite parton number.}
\eeq
This identity gives equal total rates and population derivatives for the coherent and separately resolved prescriptions at the instant the state has definite parton number. 

The coherent prescription fixes the virtual operator as
\beq\label{B20}
\hat\Sigma\,=\,\frac{1}{2}\,\hat{\mathcal R}^{\dagger}\hat{\mathcal R}
\,=\,\frac{1}{2}\sum_{k,k'\geq1}\hat{\mathcal R}^{\dagger}_{k}\hat{\mathcal R}_{k'}\,.
\eeq
Each $\hat{\mathcal R}_k$ acts on its own regulated shell, so no product $\hat{\mathcal R}_k^\dagger\hat{\mathcal R}_{k'}$ squares a distribution. The diagonal products with $k=k'$ conserve parton number and form $\Sigma_n^{\rm diag}$, which generalizes the Sudakov exponent. 
The off-diagonal products change the parton number by $k'-k$, so they can raise or lower it.
 In the component amplitude ansatz they give
\beq\label{B21}
Q^{2}\frac{\partial\psi_{n}}{\partial Q^{2}}\,=\,
-\,\Sigma^{\rm diag}_{n}\psi_{n}
\,+\,\sum_{k\geq1}\mathcal R_{n,n-k}\,\psi_{n-k}
\,-\,\frac{1}{2}\sum_{k\neq k'}\Bigl[\hat{\mathcal R}^{\dagger}_{k}\hat{\mathcal R}_{k'}\Bigr]_{n,\,n+k-k'}
\psi_{n+k-k'}\,,
\eeq
where the last term takes the input sector $n+k-k'$ to sector $n$ and therefore lowers parton number when $k>k'$.
 Its coefficient is fixed by Eq.~\eqref{B20}, and it is absent if each $k$ is treated as a separate jump.    

The contribution of a pair $(k,k')$ to an occupation derivative contains the expectation value
\beq\label{B24002}
\Bigl\langle\,\hat{\mathcal R}^{\dagger}_{k'}\bigl[N_{\lambda},\hat{\mathcal R}_{k}\bigr]\Bigr\rangle\,=\,
\sum V^{(k')*}_{b_{1}\ldots b_{k'+1},\,d}\;V^{(k)}_{a_{1}\ldots a_{k+1},\,c}\;
\Bigl(\sum_{i=1}^{k+1}\delta_{\lambda a_{i}}-\delta_{\lambda c}\Bigr)  
\Bigl\langle\,a^{\dagger}_{d}\,a_{b_{k'+1}}\cdots a_{b_{1}}\;
a^{\dagger}_{a_{1}}\cdots a^{\dagger}_{a_{k+1}}\,a_{c}\,\Bigr\rangle\,.
\eeq
For $k\neq k'$, this expectation value probes coherence between sectors differing by $k-k'$. The thermal Gaussian and mode-incoherence assumptions reduce the channel-diagonal part to
\beq\label{B24}
\frac{dn_{\lambda}}{dt}\,=\,\sum_{k\geq1}\Bigl[
\frac{k+1}{S_{V}}\!\!\sum_{\{a\},c}\!\bigl|\mathcal V^{(k)}_{\lambda a_{2}\ldots a_{k+1},c}\bigr|^{2}
n_{c}\bigl(1+n_{\lambda}\bigr)\!\!\prod_{i\geq2}\!\bigl(1+n_{a_{i}}\bigr)
\,-\,\frac{1}{S_{V}}\sum_{\{a\}}\bigl|\mathcal V^{(k)}_{a_{1}\ldots a_{k+1},\lambda}\bigr|^{2}
n_{\lambda}\prod_{i}\bigl(1+n_{a_{i}}\bigr)\Bigr]\,.
\eeq
The normalization of Eq.~\eqref{B9001} removes the repeated orderings and leaves $1/S_V$, with $S_V=(k+1)!$. The formula is bosonic. Each fermionic daughter instead contributes its Pauli factor. Coincident daughter modes require the contractions described after Eq.~\eqref{B16}.

The expectation value in Eq.~\eqref{B24002} contains
\beq\label{B24004}
\bigl(k'+2\bigr)\,+\,\bigl(k+2\bigr)\,=\,k+k'+4
\eeq
creation and annihilation operators. The count is odd for odd $k+k'$. A Gaussian state with vanishing one-point functions commutes with $(-1)^{\hat N}$. An odd product anticommutes with this parity operator and therefore has zero expectation value. This applies to thermal and Bogoliubov states, since a Bogoliubov transformation creates and destroys pairs and preserves parton-number parity. Even products can contain the anomalous covariance $m_{\lambda\mu}=\langle a_\lambda a_\mu\rangle$ in a paired Gaussian state allowed by the momentum conditions of Section~\ref{SecAWavefunctions}. The anomalous contraction also enters the number-conserving case $k=k'$, for example $\langle a^\dagger a^\dagger aa\rangle=2n^2+|m|^2$. Equation~\eqref{B24} requires the thermal Gaussian case without anomalous covariance.

The $1\to2$ and $1\to3$ interference has odd $k+k'$ and vanishes with the stated Gaussian assumption. The $1\to2$ and $1\to4$ interference has even $k+k'$ and can receive paired Gaussian contributions containing $m$ only when the state permits nonzero anomalous covariance. Those contributions vanish for the thermal state.  
A coherent classical field can give a nonzero one-point function in a fixed color background, as in the Color Glass Condensate of Ref.~\cite{ArmestoCGC}.\footnote{The operator $a_\lambda$ carries color, so its one-point function vanishes in a color-singlet hadron by color conservation. 
The static classical sources in the Color Glass Condensate can absorb longitudinal momentum while their recoil is neglected.
 Their fixed-background field therefore does not obey the momentum constraint of an isolated hadron state. Averaging the sources removes a colored one-point function. A color-singlet pair is allowed by color alone, but the momentum condition of Section~\ref{SecAWavefunctions} still forbids $\langle a_\lambda a_\mu\rangle$ for a state commuting with $\hat P^+$. Color neutrality alone does not imply Gaussianity or parton-number parity.}

Momentum conservation at every vertex gives
\beq\label{B19005}
[\hat P^{+},\hat{\mathcal R}_{k}]=0
\eeq
which also holds for a merging vertex with the same momentum delta functions. Momentum conservation therefore constrains such a vertex although norm conservation does not fix its normalization. Equation~\eqref{eq:appc-4} gives the operator argument for off-diagonal channel pairs and arbitrary states.

\subsection{Loop counting}

The coefficient $P^{(p)}$ multiplies $(\alpha_s/2\pi)^{p+1}$ in the expansion of $P_{ij}$, in the convention of Eq.~\eqref{B1002}.
Each splitting amplitude has its own loop expansion,
\beq\label{B26}
\hat{\mathcal R}_{k}\,=\,\sum_{\ell\geq0}\hat{\mathcal R}^{(\ell)}_{k}\,,\qquad
\hat{\mathcal R}^{(\ell)}_{k}\,\sim\,g_{s}^{\,k}\,\alpha_{s}^{\,\ell}\,,
\eeq
and the product of the $\ell$-loop and $\ell'$-loop vertices with the same final state contributes to $P^{(p)}$ according to
\beq\label{B27}
\hat{\mathcal R}^{(\ell)\dagger}_{k}\,\hat{\mathcal R}^{(\ell')}_{k}\,\sim\,\alpha_{s}^{\,k+\ell+\ell'}\,,\qquad
k\,-\,1\,+\,\ell\,+\,\ell'\,=\,p\,,
\eeq
which at $p=2$ includes the square of the one-loop vertex. The tree $1\to3$ contribution enters $P^{(1)}$ through its own square, with the strongly ordered region of two successive $1\to2$ splittings subtracted. The counting equations alone supply neither an inclusive kernel nor the phases of its amplitudes.

\subsection{Paired Gaussian covariance}

The paired Gaussian formulas below require the packet or noncommuting-state conditions of Section~\ref{SecAWavefunctions}.
 Such a Gaussian state requires both the normal covariance $n_{\lambda\mu}=\langle a^\dagger_\mu a_\lambda\rangle$ and the anomalous covariance $m_{\lambda\mu}=\langle a_\lambda a_\mu\rangle$. We take $\langle a_\lambda\rangle=0$. If the mean is nonzero, $n$ and $m$ in the covariance formula are the connected covariances. For a physical bosonic covariance, the symplectic eigenvalues give the entropy of the Gaussian state $\hat\rho_G$,
\beq\label{B29}
\nu_{\lambda}\,=\,{\rm eig}_{+}
\begin{pmatrix} n+\frac{1}{2} & m \\ -m^{\dagger} & -n^{T}-\frac{1}{2}\end{pmatrix},\qquad
S_{\rm vN}(\hat\rho_{G})\,=\,\sum_{\lambda}\Bigl[\Le\nu_{\lambda}+\frac{1}{2}\Ra\ln\Le\nu_{\lambda}+\frac{1}{2}\Ra
-\Le\nu_{\lambda}-\frac{1}{2}\Ra\ln\Le\nu_{\lambda}-\frac{1}{2}\Ra\Bigr]\,.
\eeq
The anomalous matrix need not be diagonal in the natural-orbital basis of $n$. A two-mode squeezed state has diagonal $n$, off-diagonal $m$ and zero entropy, but for positive longitudinal fractions it does not commute with $\hat P^+$. If $m$ is diagonal in that basis, the eigenvalues reduce to
\beq\label{B29001}
\nu_{\lambda}\,=\,\sqrt{\Le n_{\lambda}+\frac{1}{2}\Ra^{2}\,-\,\bigl|m_{\lambda}\bigr|^{2}}\,.
\eeq
The physical covariance condition is $|m_\lambda|^2\leq n_\lambda(n_\lambda+1)$. At $m=0$, the entropy reduces to $S_{\rm mode}$ of Eq.~\eqref{AB20}. At fixed normal covariance, pairing lowers the Gaussian entropy relative to the thermal state. This comparison does not show that the splitting dynamics preserves a Gaussian state. Berges, Floerchinger and Venugopalan~\cite{BFV} use the Gaussian entropy with anomalous correlations for expanding quantum fields.

\end{document}